\documentclass[aoas]{imsart}

\RequirePackage{amsthm,amsmath,amsfonts,amssymb}
\RequirePackage[authoryear]{natbib}
\RequirePackage{mathrsfs}
\RequirePackage{xcolor}
\RequirePackage{dsfont}
\RequirePackage[colorlinks,citecolor=blue,urlcolor=blue]{hyperref}
\RequirePackage{graphicx}
\usepackage{booktabs}

\startlocaldefs
\theoremstyle{plain}
\newtheorem{theorem}{Theorem}
\theoremstyle{definition}

\def\T{{\mathrm{\scriptscriptstyle T}}}

\newcommand{\mbf}[1]{\boldsymbol{#1}}

\endlocaldefs

\begin{document}

\begin{frontmatter}
\title{Semiparametric Functional Multistate Modeling of Alzheimer’s Disease Progression with Imaging Biomarkers}
\runtitle{Functional Multistate Modeling of AD Progression}

\begin{aug}
\author[A]{\fnms{Chenrui}~\snm{Qi}\ead[label=e1]{qichenrui@connect.hku.hk}}
\author[B]{\fnms{Kai}~\snm{Kang}\ead[label=e2]{kangk5@mail.sysu.edu.cn}}
\author[A]{\fnms{Yu}~\snm{Gu}\ead[label=e3]{yugu@hku.hk}}
\address[A]{Department of Statistics and Actuarial Science, University of Hong Kong, Pokfulam Road, Hong Kong\printead[presep={,\ }]{e1,e3}}
\address[B]{Department of Statistics, Sun Yat-sen University, Guangzhou, China\printead[presep={,\ }]{e2}}
\end{aug}

\begin{abstract}
Medical imaging provides rich information for predicting Alzheimer’s disease progression, but existing imaging-based methods typically focus on a single survival endpoint and treat transition times as exactly observed or right-censored. Motivated by the Alzheimer's Disease Neuroimaging Initiative (ADNI), we develop a predictive framework that represents disease progression as an intermittently observed multistate process with interval-censored transition times and predicts future progression from any current disease state. We incorporate imaging biomarkers as functional covariates in a semiparametric proportional intensity model and combine functional principal component analysis with nonparametric maximum pseudo likelihood estimation. We further develop a profile score test for assessing the overall association between the imaging covariate and the multistate process. We establish the asymptotic properties of the proposed estimators and test statistic, and simulation studies demonstrate satisfactory finite-sample performance. In the ADNI application, baseline lateral ventricular morphology is strongly associated with Alzheimer’s disease progression. The proposed functional multistate model also achieves the best overall predictive performance among the competing methods.
\end{abstract}


\end{frontmatter}

\newcommand{\SuppSecEMEquiv}{S.1.1}
\newcommand{\SuppSecEMCondExp}{S.1.2}
\newcommand{\SuppSecConditions}{S.2.1}
\newcommand{\SuppSecProofThmOne}{S.2.3}
\newcommand{\SuppSecProofThmTwo}{S.2.4}
\newcommand{\SuppSecMRI}{S.3}

\newcommand{\SuppCondOne}{1}
\newcommand{\SuppCondTen}{10}

\newcommand{\SuppFigCumHaz}{S1}
\newcommand{\SuppFigTransProb}{S2}

\section{Introduction}
\label{s:intro}
According to the World Health Organization, approximately 57 million people worldwide were living with dementia in 2021, with nearly 10 million new cases occurring each year. Alzheimer's disease (AD) is the most common form of dementia and may account for 60--70\% of cases \citep{who2026dementia}.
Current consensus criteria view AD as a biological process that may begin years before overt symptoms, with its clinical manifestations subsequently unfolding along a continuum of cognitive and functional decline \citep{scheltens2021alzheimer,jack2024revised}. 
Clinically, AD progression is often characterized by three disease states: cognitively normal (CN), mild cognitive impairment (MCI), and AD \citep{petersen2004mild,petersen2014mild}.
However, there is substantial heterogeneity in both the timing and pace of this progression. For example, MCI is a clinically important intermediate state in which some individuals remain stable for extended periods, whereas others progress rapidly to AD. Identifying biomarkers that predict both the early transition from CN to MCI and the later transition from MCI to AD is therefore important for individualized risk assessment and accurate disease prediction.

Among candidate biomarkers, medical imaging provides rich information for characterizing disease progression and predicting individual risk trajectories. For example, structural magnetic resonance imaging (MRI) measures of regions such as the hippocampus are strongly associated with cognitive decline and AD progression \citep{frisoni2010clinical}. 
In an early Alzheimer's Disease Neuroimaging Initiative (ADNI) investigation, ventricular enlargement differs across CN, MCI, and AD groups and is greater among individuals with MCI who subsequently progress to AD, supporting its value as a marker of disease progression \citep{nestor2008ventricular}. More generally, surface-based analyses can reveal localized morphometric changes that are obscured by a single volumetric summary \citep{dong2020applying}. Similarly, voxel-level imaging measurements can preserve spatial information that may be lost when images are reduced to summary measures. Because these measurements are densely sampled across spatial locations and exhibit strong local correlation, it is more natural to preserve their spatial structure and treat them as functional covariates than to reduce them to a few prespecified regional summaries.
Existing work linking imaging biomarkers to AD risk mostly falls within a single-event survival framework. Baseline images have been treated as functional covariates in Cox regression models \citep{lee2015bflcrm,kong2018flcrm}, and functional joint models have been developed to connect longitudinal clinical outcomes with survival, with imaging data entering as baseline functional predictors \citep{li2017functional,wang2020partial,zou2021bayesian} or longitudinal functional outcomes \citep{li2019bayesian,kang2023joint,zhou2024joint,zou2025dynamic}. 
These methods have demonstrated the prognostic value of imaging biomarkers for AD conversion. However, they all focus on a single survival outcome, typically progression from MCI to AD, and thus restrict their analyses to participants with MCI at baseline. Moreover, they all treat the time to AD onset as exactly observed or right-censored. In longitudinal AD studies such as ADNI, however, disease state is assessed only at scheduled visits, so each transition is known only to occur between two successive visits and is therefore interval-censored \citep{petersen2010}. As a result, these approaches do not fully use information from participants who are cognitively normal at enrollment, do not reflect uncertainty about the transition time between visits, and when transitions are modeled separately, fail to account for their within-subject dependence. This dependence is important because the earlier CN-to-MCI transition may provide valuable information for predicting subsequent progression from MCI to AD.

Multistate models provide a natural framework for modeling the full sequence of intermittently observed disease states \citep{kalbfleisch1985analysis,jackson2011multistate,titman2011flexible,gu2024maximum}. However, these methods are not designed for high-dimensional imaging biomarkers. Directly entering hundreds or thousands of image measurements as ordinary covariates is statistically unstable and ignores their spatial structure. Incorporating a functional imaging covariate is also technically difficult because the interval-censored multistate likelihood must integrate over subject-specific random effects that capture dependence between transitions and sum over unobserved transition paths, while the infinite-dimensional imaging effect must be estimated simultaneously.

To fill these gaps, we propose a semiparametric functional multistate model to predict the entire disease trajectory using imaging biomarkers and interval-censored multistate data. We use functional principal component analysis (FPCA) for dimension reduction, which transforms the original functional covariates into a diverging number of scalar covariates. We then adopt a Poisson data augmentation strategy and perform nonparametric maximum pseudo likelihood estimation via an efficient EM algorithm. Furthermore, we develop a profile-likelihood-based score test to evaluate the association between imaging covariates and the multistate disease process.
We establish the consistency and convergence rates of the proposed estimators and derive the asymptotic null distribution of the test statistic, addressing the major technical challenges introduced by FPCA. 
Finally, we assess the performance of the proposed methods in realistic settings through comprehensive simulation studies and an application to the Alzheimer's Disease Neuroimaging Initiative.
The ADNI analysis reveals a strong association between baseline lateral ventricular morphology and disease progression, while the proposed functional multistate model achieves the best overall predictive performance for MCI-to-AD progression among the competing methods.

To our knowledge, this is the first work on multistate models with functional imaging covariates. Compared with existing approaches based on a single right-censored survival outcome, our multistate modeling framework has several advantages. First, it makes full use of all available data and provides predictions for all subjects, regardless of their current disease states.
Second, by modeling the entire disease trajectory as a multistate process, our approach accounts for population heterogeneity in progression rates, thereby enabling more accurate and individualized predictions. Third, our method is specifically designed for interval-censored multistate data, making it particularly well suited to real-world studies.


\section{Motivation}
\label{s:motivation}

ADNI is a longitudinal, multicenter observational study designed to identify and validate biomarkers for early diagnosis, monitor disease progression, and facilitate the development of effective treatments for AD \citep{petersen2010}. At each visit (approximately every six or twelve months), participants undergo comprehensive clinical, cognitive, neuroimaging, and biological assessments and are classified as CN, MCI, or AD. We use the 555 participants included in the final analysis to motivate our modeling framework; full cohort construction and exclusion criteria are described in Section~\ref{subsec:adni}. This cohort consists of participants who are CN or MCI at baseline and whose disease states are observed intermittently over a mean follow-up of 5.4 years.
Our primary scientific interests are twofold: to investigate the association between baseline lateral ventricular morphology and AD progression, and to predict each participant's future disease trajectory.

\subsection{Observed disease trajectories and interval censoring}
\label{subsec:adni_trajectories}
\begin{figure}
    \centering
    \includegraphics[width=1.0\linewidth]{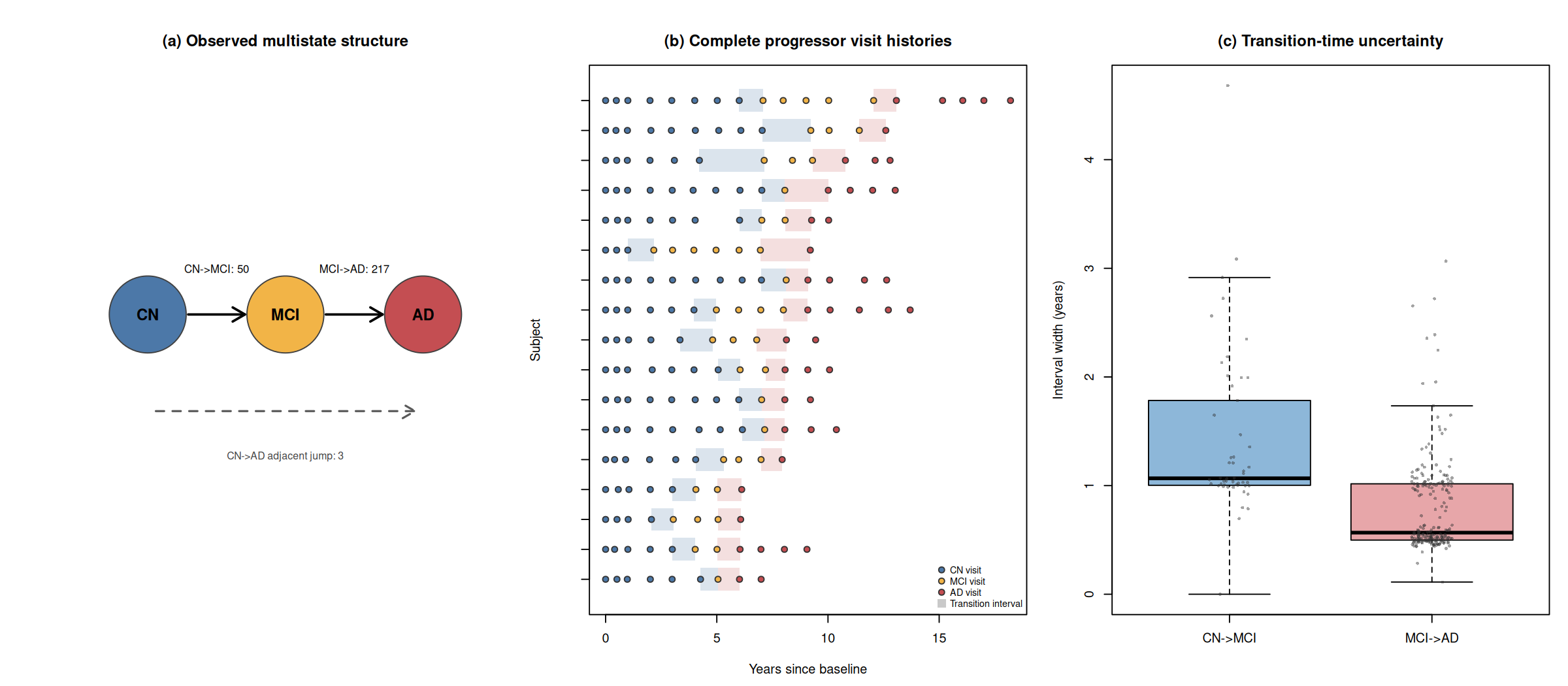}
    \caption{Observed disease histories in the ADNI ventricle model-fitting subset.
    Panel (a) shows the modeled transition structure and the observed adjacent state changes. Panel (b) shows visit histories for the 17 participants with the fully observed path CN-to-MCI-to-AD; points are examination times and shaded bands are intervals known to contain a transition. Panel (c) shows the widths of adjacent visit intervals containing CN-to-MCI and MCI-to-AD changes.}
    \label{fig:motivation_trajectory}
\end{figure}

The ADNI diagnosis records naturally define a progressive multistate process rather than a single terminal endpoint. We observe 50 adjacent CN-to-MCI changes and 217 adjacent MCI-to-AD changes, indicating that both cognitive decline and later progression to AD contribute information; see Figure~\ref{fig:motivation_trajectory} (a).
A conventional survival analysis of MCI-to-AD progression would use only the later part of the disease process. In particular, reducing the data to a single MCI-to-AD endpoint would exclude the 205 participants who are CN at baseline and ignore the 50 observed CN-to-MCI changes. It would therefore lose information about early progression and would not support prediction from the CN state.

In addition, as shown in Figure~\ref{fig:motivation_trajectory} (a), three adjacent visit pairs move directly from CN to AD. These records should not be interpreted as evidence that the biological disease process skips MCI; rather, they indicate that the intermediate MCI state must have occurred between two adjacent visits. This further motivates an interval-censored multistate likelihood that accounts for latent transition paths between visits. 

Figure~\ref{fig:motivation_trajectory} (b)--(c) illustrate why transition times cannot be treated as exactly observed.
If a participant is CN at one visit and MCI at the next, the transition is only known to have occurred between these two visits. Such observed transition intervals are often several months to more than one year wide. Thus, assigning each transition to a single time point, such as the first visit at which the new diagnosis is observed, would ignore substantial uncertainty about the true transition time.

\subsection{Functional imaging patterns and model motivation}
\label{subsec:adni_imaging_motivation}
ADNI provides baseline structural MRI measurements of lateral ventricular morphology. Following standardized image registration and segmentation, these measurements are organized as a one-dimensional functional covariate over 408 spatially ordered voxel locations; preprocessing details are provided in Section~\ref{MRI_preprocessing}. We further represent each participant's ventricular profile by a small number of functional principal component (FPC) scores using FPCA; the formal construction is given in Section~\ref{subsec:fpca}.
The ADNI data suggest that the same pattern of ventricular morphology may carry different prognostic information at different disease stages. To explore this possibility, we compare baseline ventricular FPC scores between participants who progress within 36 months and those who remain stable for at least 36 months, separately within the CN and MCI risk sets. For each retained FPC score, we summarize the contrast by the standardized mean difference, defined as the difference between the progressor and stable group means divided by the pooled standard deviation. This scale-free summary allows contrasts from different FPC scores to be compared on the same axis.

\begin{figure}
    \centering
    \includegraphics[width=1.0\linewidth]{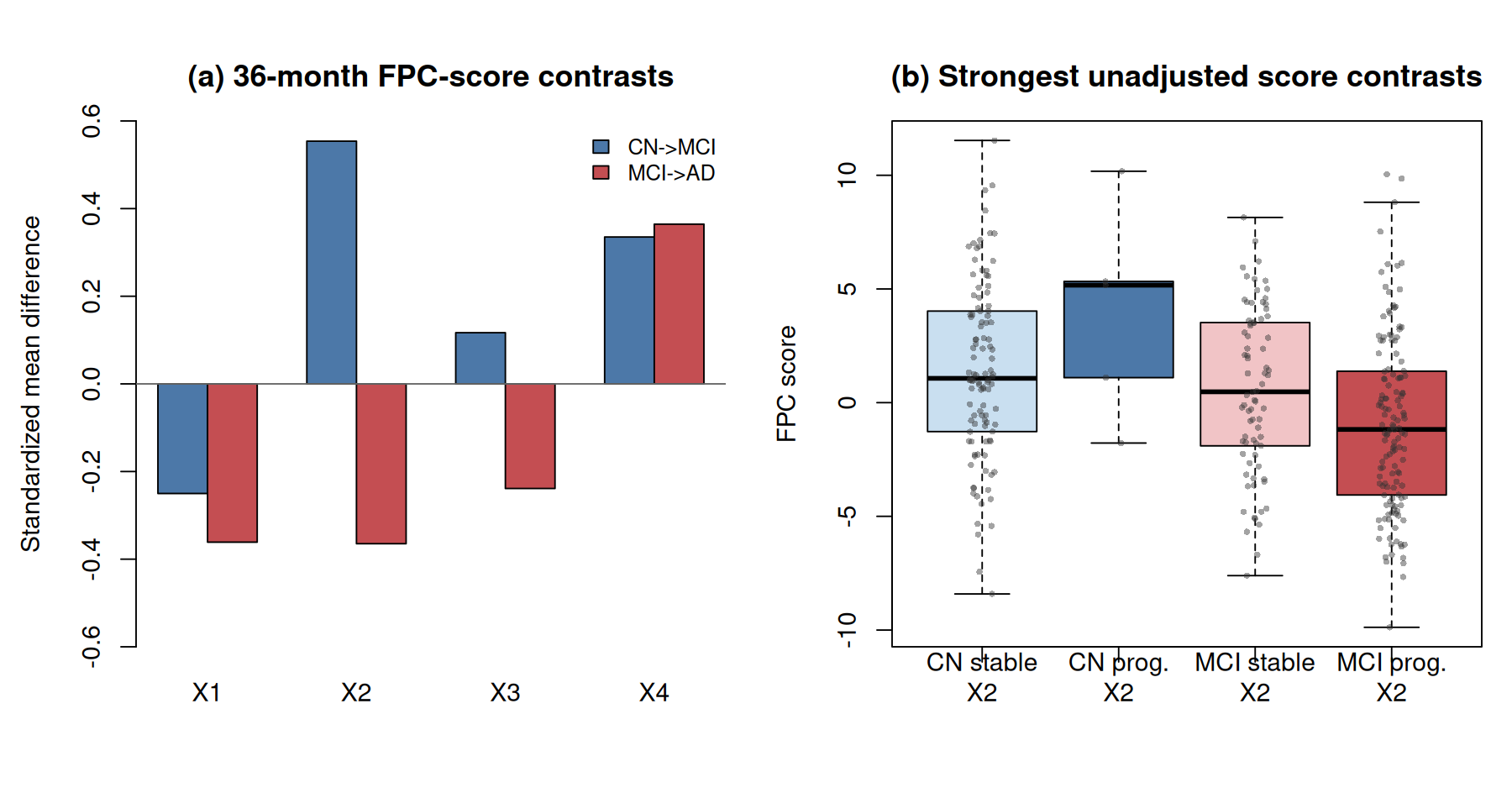}
    \caption{Exploratory transition-specific contrasts of baseline ventricular FPC scores in the ADNI ventricle model-fitting subset. Progressors reach the target state within 36 months; stable participants have at least 36 months of follow-up without the target transition.}
    \label{fig:motivation_fpc_contrasts}
\end{figure}

Figure~\ref{fig:motivation_fpc_contrasts} shows that the exploratory FPC score contrasts are not identical for the two transitions. For example, the second FPC score has a positive contrast for CN-to-MCI progression but a negative contrast
for MCI-to-AD progression, whereas the fourth FPC score shows positive contrasts for both transitions. The right panel displays the strongest unadjusted score contrast, illustrating that the same FPC score can separate progressors from stable participants in opposite directions across the two risk sets. Because only a small number of CN participants progress within 36 months, these contrasts should be interpreted descriptively rather than as formal evidence for a particular FPC effect. Nevertheless, they suggest that ventricular morphology may carry transition-specific prognostic information.

These observations motivate a multistate model that incorporates ventricular morphology as a functional covariate with transition-specific coefficients. We further include a shared subject-specific random effect to account for dependence between transitions arising from unobserved progression characteristics.

\section{Methods}
\label{s:methods}

\subsection{Model setup}
\label{subsec:model_setup}
Motivated by this exploration of ADNI patterns, we formulate a semiparametric model for a general multistate process with $K$ disease states. 
Let $\mathcal{D}$ denote the set of all distinct state pairs $(j,k)$ such that the transition from state $j$ to state $k$ is feasible.
In the ADNI application, states 1, 2, and 3 denote CN, MCI, and AD, respectively, and $\mathcal{D}=\{(1,2), (2,3)\}$. 
For the $i$th subject ($i=1,\dots,n$), let $\mbf{X}_i(t)$ denote a $d$-vector of potentially time-dependent covariates and $b_i\sim N(0,\sigma^2)$ denote the latent Gaussian random effect with unknown variance $\sigma^2$. 
In addition, let $\{M_i(v): v\in\mathcal{V}\}$ denote the functional imaging covariate converted from the $i$th subject's original image, where $\mathcal{V}$ is a dense domain with each element $v\in\mathcal{V}$ corresponding to a voxel in 3D images (or a pixel in 2D images). In the ADNI application, $\mathcal{V}$ is the common one-dimensional domain of 408 spatially ordered voxel locations within the lateral ventricle region of interest (ROI), $v$ indexes a location in this ordering, and $M_i(v)$ represents the corresponding imaging measurement. 

We relate the $i$th subject's transition intensity functions to the covariates $\mbf{X}_i(t)$ and $M_i(v)$, and the random effect $b_i$ through a functional proportional intensity model:
\begin{equation} \label{eq1}
    h_{ijk}(t \mid \mbf{X}_i, M_i, b_i) = h_{jk}(t)\exp\left\{\mbf{\gamma}_{jk}^{\T}\mbf{X}_i(t) +\int_{\mathcal{V}}\beta_{jk}(v)M_i(v)dv+b_i\right\}, \quad \text{ for } (j,k)\in\mathcal{D},
\end{equation}
where $h_{jk}(t)$ is an arbitrary baseline intensity function, $\mbf{\gamma}_{jk}$ is a $d$-vector of unknown scalar regression coefficients, and $\beta_{jk}(v)$ is an unknown functional regression coefficient. Here, the scalar random effect $b_i$ is shared across all transitions $(j,k)\in\mathcal{D}$ to account for their within-subject dependence. This parsimonious specification is appropriate for our ADNI application, which contains only two possible transitions and only 17 participants observed to experience both CN-to-MCI and MCI-to-AD transitions during follow-up. For more general multistate processes with more complex dependence structures, model~\eqref{eq1} can be extended to include multi-dimensional Gaussian random effects through their interactions with covariates, and the estimation procedure described below can be adapted accordingly.

\subsection{Functional principal component analysis}
\label{subsec:fpca}
Model~\eqref{eq1} involves the functional coefficients $\beta_{jk}(v)$ and functional covariate $M_i(v)$, which are all infinite-dimensional and cannot be incorporated into estimation directly. 
We therefore use FPCA to obtain parsimonious representations of these functional terms. 

Let $\mu(v) = E\left\{M_i(v) \right\}$ and $\Sigma(v,v^{\prime}) = \text{cov}\{M_{i}(v),M_{i}(v^{\prime})\}$ denote the mean and covariance functions of $M_i(v)$, respectively.
By Mercer's theorem, $\Sigma(v,v^{\prime})$ admits the spectral decomposition  $\Sigma(v,v^{\prime})=\sum_{l=1}^{\infty}\lambda_{l}\phi_{l}(v)\phi_{l}(v^{\prime})$,
where $\{\lambda_l\}_{l=1}^{\infty}$ is a non-increasing sequence of nonnegative eigenvalues and $\{\phi_{l}(v)\}_{l=1}^{\infty}$ are the corresponding orthonormal eigenfunctions.
It follows from the Karhunen-Lo\`eve expansion that $$M_i(v)=\mu(v)+\sum_{l=1}^{\infty}\xi_{il}\phi_{l}(v), \qquad \xi_{il}=\int_{\mathcal{V}}\{M_{i}(v)-\mu(v)\}\phi_{l}(v)dv$$
where the FPC scores $ \xi_{il}$ summarize individual deviations from $\mu(v)$ in the directions of the eigenfunctions $\phi_l(v)$.

We then project $\beta_{jk}(v)$ onto the span of the eigenfunctions $\{\phi_l(v)\}_{l=1}^{\infty}$ to obtain $\beta_{jk}(v)=\sum_{l=1}^\infty\beta_{jkl}\phi_l(v)$,
where $\beta_{jkl} = \int_{\mathcal{V}} \beta_{jk}(v)\phi_l(v)dv$.
Substituting these expansions into model \eqref{eq1} yields
\begin{equation}
        h_{ijk}(t \mid \mbf{X}_i, M_i, b_i) =\check{h}_{jk}(t)\exp\left\{\mbf{\gamma}_{jk}^\T\mbf{X}_i(t)+ \sum_{l=1}^{\infty}\beta_{jkl}\xi_{il}+b_i\right\}, \quad \text{ for } (j,k)\in\mathcal{D},
        \label{eq2}
\end{equation}
where $\check{h}_{jk}(t) = h_{jk}(t) \exp \left\{\int_{\mathcal{V}}\beta_{jk}(v)\mu(v)dv\right\}$.
Let $\hat{\mu}(v)$, $\hat{\phi}_l(v)$ and  $\hat \xi_{il}$ denote the estimated mean function, eigenfunctions, and FPC scores, respectively, obtained using standard FPCA techniques \citep{ramsay2005functional}.
In practice, we substitute $\xi_{il}$ with $\hat \xi_{il}$ and truncate the infinite sum at a finite number $r_n$, which grows with the sample size $n$. Then, model \eqref{eq2} reduces to a proportional intensity model with a diverging number of covariates:
\begin{equation}
    h_{ijk}(t \mid \mbf{X}_i, M_i, b_i) =\check{h}_{jk}(t)\exp\left\{\mbf{\gamma}_{jk}^{\T}\mbf{X}_{i}(t)+
    \mbf{B}_{jk}^{\T}\hat{\mbf{\Xi}}_i + b_i\right\}, \quad \text{ for } (j,k)\in\mathcal{D}.
    \label{eq3}
\end{equation}
where $\mbf{B}_{jk} = (\beta_{jk1}, \ldots, \beta_{jkr_n})^{\T}$ and $\hat{\mbf{\Xi}}_i = ({\hat\xi}_{i1}, \ldots, {\hat\xi}_{ir_n})^{\T}$.
Model \eqref{eq3} involves three sets of quantities to be estimated: (i) the scalar regression parameters $\mbf{\theta} = [\{\mbf{\gamma}_{jk}\}_{(j,k) \in \mathcal{D}}, \sigma^2 ]$, (ii) the FPCA-truncated imaging effects $\mathcal{B}_{r_n} = \{\mbf{B}_{jk}\}_{(j,k) \in \mathcal{D}}$ , and (iii) the cumulative baseline intensity functions $\mathcal{H}=\{H_{jk} \}_{(j,k) \in \mathcal{D}}$, with $ H_{jk}(t)=\int_0^t h_{jk}(s)ds$.

\subsection{Intermittent observation and pseudo likelihood}
\label{subsec:data_likelihood}
To estimate $\mbf{\theta}$, $\mathcal{B}_{r_n}$, and $\mathcal{H}$ jointly, we first need to express the likelihood implied by model~\eqref{eq3} in terms of the observed data.
A key challenge in studying AD is that cognitive progression operates as a continuous-time process, whereas disease states are observed only at discrete clinical visits. This results in the following general interval-censoring scheme. 
For subject \(i\), let
$0=\tau_{i0} < \tau_{i1} < \cdots < \tau_{iQ_i}$
denote the $(Q_i +1)$ examination times and let $(S_{i0}, S_{i1}, \ldots, S_{iQ_i})$  denote the corresponding states observed at each examination time. The observed data for subject $i$ is $$O_i = \left\{(\tau_{i0},\ldots,\tau_{iQ_i}),(S_{i0},\ldots,S_{iQ_i}),\mbf{X}_i(\cdot), M_i(\cdot)\right\}.$$
Given the protocol-based ADNI visit schedule, we impose the independent censoring assumption that the density of $\tau_{iq}$ ($q=1,\dots,Q_i$) at time $t$ depends only on the observed history up to the current time, including $\{\mbf{X}_i(s): s\le t\}$, $M_i(v)$, and $\{(\tau_{iq'}, S_{iq'}): q'=0,\dots,q-1\}$. 

To formulate the pseudo likelihood under model~\eqref{eq3}, we define the $K\times K$ cumulative transition intensity matrix $\mbf{A}_i(t;\mbf{X}_i,\hat{\mbf{\Xi}}_i,b_i)$, whose
$(j,k)$th off-diagonal element is
$
A_i^{(j,k)}(t;\mbf{X}_i,\hat{\mbf{\Xi}}_i,b_i)
=
\int_0^t
\exp\{
\mbf{\gamma}_{jk}^{\T}\mbf{X}_i(s)+
\mbf{B}_{jk}^{\T}\hat{\mbf{\Xi}}_i+b_i
\}
\,d\check H_{jk}(s) 
$ if $(j,k)\in\mathcal{D}$ 
and $0$ otherwise.
The diagonal elements are defined by
$
A_i^{(j,j)}(t;\mbf{X}_i,\hat{\mbf{\Xi}}_i,b_i)
=
-
\sum_{k\neq j}
A_i^{(j,k)}(t;\mbf{X}_i,\hat{\mbf{\Xi}}_i,b_i).
$
Because latent transitions may occur between examinations, transition probabilities are obtained through the product integral representation \cite[Theorem II.6.7]{andersen1993statistical}. 
Specifically, for any $0\le t_1\le t_2$, the $K\times K$ transition probability matrix is given by
$$
\mbf{P}_i(t_1,t_2;\mbf{X}_i,\hat{\mbf{\Xi}}_i,b_i)
=
\pi_{t_1}^{t_2}
\left\{
\mbf{I}_K+d\mbf{A}_i(s;\mbf{X}_i,\hat{\mbf{\Xi}}_i,b_i)
\right\},
$$
where
$
\pi_{t_1}^{t_2}\{\mbf{I}_K+d\mbf{A}_i(s)\}
=
\lim_{\max_l|s_l-s_{l-1}|\to 0}
\prod_{l=1}^{L}
\left\{
\mbf{I}_K+\mbf{A}_i(s_l)-\mbf{A}_i(s_{l-1})
\right\},
$
for a partition $t_1=s_0<s_1<\cdots<s_L=t_2$ of $(t_1,t_2]$, and $\mbf{I}_K$ is the
$K\times K$ identity matrix. 

Under the independent censoring assumption and the conditional Markov assumption given the random effects $b_i$, the conditional pseudo likelihood given the initial states is
\begin{equation}
    \tilde L_n(\mbf{\theta},\mathcal{B}_{r_n},\mathcal{H})= \prod_{i=1}^n\int_{b_i}\prod_{q=1}^{Q_i}\mbf{P}_i(\tau_{i,q-1},\tau_{iq};\mbf{X}_{i},\hat{\mbf{\Xi}}_i,b_{i})^{(S_{i,q-1},S_{iq})} f(b_i;\sigma^2)db_i,
    \label{eq4}
\end{equation}
where $\mbf{P}^{(j,k)}$ denotes the $(j,k)$th element of the matrix $\mbf{P}$ and $f(b;\sigma^2) = (2\pi\sigma^2)^{-1/2}\exp\{-b^2/(2\sigma^2)\}$. We call $\tilde L_n(\mbf{\theta},\mathcal{B}_{r_n},\mathcal{H})$ pseudo likelihood because it is an approximation of the true likelihood by using only the first $r_n$ estimated FPC scores $\hat{\mbf{\Xi}}_i$ to represent the full imaging covariates.

\subsection{Nonparametric maximum pseudo likelihood estimation}
\label{subsec:npmple}
Direct maximization of the observed-data pseudo likelihood $\tilde L_n(\mbf{\theta},\mathcal{B}_{r_n},\mathcal{H})$ is computationally intractable due to the infinite-dimensional baseline intensity functions $\mathcal{H}$. To address this, we extend the nonparametric maximum likelihood estimation approach to estimate each component of $\mathcal{H}$.
Specifically, for $(j,k)\in\mathcal{D}$, let $\check{H}_{jk}(t) = \int_0^t \check{h}_{jk}(s)ds$ be the transformed cumulative baseline intensity function.
We treat each $\check{H}_{jk}$ as a step function with nonnegative jumps at all examination times $\tau_{iq}$ ($i=1,\dots,n$ and $q=1,\dots,Q_i$). 

Even after this reduction, however, the resulting jump parameters have no closed-form expressions: each subject's contribution to $\tilde L_n$ involves a product integral as well as an integration over the random effect $b_i$, so maximizing $\tilde L_n(\mbf{\theta},\mathcal{B}_{r_n},\mathcal{H})$ with respect to $\mbf{\theta}$, $\mathcal{B}_{r_n}$, and the jump sizes of each $\check{H}_{jk}$ remains infeasible. To overcome this second difficulty, we use a Poisson data augmentation method and maximize the pseudo likelihood via an EM algorithm.
Specifically, let $0<u_1<\cdots<u_m$ be the ordered distinct examination times pooled across all subjects, and let $
\Delta \check H_{jk}(u_s)=\check H_{jk}(u_s)-\check H_{jk}(u_{s-1})$
denote the jump size of $\check H_{jk}$ at $u_s$. For each subject $i$, transition $(j,k)\in\mathcal{D}$, and time point $u_s$, we introduce a latent Poisson random variable $W_{ijks}$ with mean $ \Delta \check H_{jk}(u_s) \exp\{ \mbf{\gamma}_{jk}^{\T}\mbf{X}_i(u_s)+ \mbf{B}_{jk}^{\T}\hat{\mbf{\Xi}}_i + b_i\}$. Conceptually, $W_{ijks}$ represents the unobserved number of transitions from state $j$ to state $k$ experienced by subject $i$ over the time interval $(u_{s-1}, u_s]$.
As shown in Section~\SuppSecEMEquiv\ of the Supplementary Material, these latent Poisson variables induce a pseudo likelihood that is equivalent to the original pseudo likelihood in \eqref{eq4}. Thus, it suffices to maximize the Poisson-based pseudo likelihood via an EM algorithm, by treating these Poisson variables $W_{ijks}$ ($i=1,\ldots,n; (j,k) \in \mathcal{D}; s=1,\ldots,m$) and the random effects $b_i$ ($i=1,\ldots,n$) as missing data.
The complete-data pseudo log-likelihood is expressed as: 
\begin{align*}
\sum_{i=1}^n \Bigg[
&\sum_{(j,k)\in\mathcal{D}} \sum_{s=1}^m \mathds{1}(u_s \le \tau_{i,Q_i})
\Bigg\{ 
W_{ijks}
\Big[
\log\{\Delta \check H_{jk}(u_s)\}
+\mbf{\gamma}_{jk}^{\T}\mbf{X}_i(u_s) +\mbf{B}_{jk}^{\T}\hat{\mbf{\Xi}}_i
+b_i
\Big] \\
&\qquad\qquad\qquad\qquad
-
\Delta \check H_{jk}(u_s)
\exp\left\{
\mbf{\gamma}_{jk}^{\T}\mbf{X}_i(u_s)
+\mbf{B}_{jk}^{\T}\hat{\mbf{\Xi}}_i
+b_i
\right\} -\log(W_{ijks}!)
\Bigg\} \\
&\qquad\qquad\qquad\qquad
-\frac{1}{2}\log(2\pi)-\log\sigma-\frac{b_i^2}{2\sigma^2}
\Bigg],
\end{align*}
where $\mathds{1}(\cdot)$ is the indicator function.

Let $\widetilde E(\cdot)$ denote the conditional expectation given the observed data $O_i$. In the E-step, we evaluate the conditional expectations $\widetilde{E}(W_{ijks})$, $\widetilde E\{\exp(b_i)\}$, and $ \widetilde E(b_i^2)$. The conditional density of $b_i$ given $O_i$, denoted by $f(b_i\mid O_i)$, is proportional to $$\prod_{q=1}^{Q_i}\mbf{P}_i(\tau_{i,q-1},\tau_{iq};\mbf{X}_i,\hat{\mbf{\Xi}}_i,b_i)^{(S_{i,q-1},S_{iq})}f(b_i;\sigma^2).$$ The conditional expectation of $W_{ijks}$ given $O_i$ can be computed as 
$ \widetilde E(W_{ijks})=\int_{b_i} E(W_{ijks}\mid O_i,b_i)\,f(b_i\mid O_i)\,db_i$,
with the derivation of $E(W_{ijks}\mid O_i,b_i)$ detailed in Section~\SuppSecEMCondExp\ of the Supplementary Material. Integrals over $b_i$ can be approximated using Gauss--Hermite quadratures. 

In the M-step, the jump size of $\check H_{jk}$ at $u_s$ is updated in closed form by 
\[
\Delta \check H_{jk}^{\,\mathrm{new}}(u_s)
=
\frac{
\sum_{i=1}^n \mathds{1}(u_s\le \tau_{i,Q_i})\,\widetilde E(W_{ijks})
}{
\sum_{i=1}^n \mathds{1}(u_s\le \tau_{i,Q_i})
\exp\bigl\{\mbf{\gamma}_{jk}^{\T}\mbf{X}_i(u_s)
+\mbf{B}_{jk}^{\T}\hat{\mbf{\Xi}}_i\bigr\}\widetilde E\{\exp(b_i)\}
},
\]
for $(j,k)\in\mathcal D$ and $s=1,\ldots,m$.
After substituting the updated jump sizes $\Delta \check H_{jk}^{\,\mathrm{new}}(u_s)$ into the expected complete-data pseudo log-likelihood, we update $(\mbf{\gamma}_{jk},\mbf{B}_{jk})$ for each $(j,k)\in\mathcal D$ by solving the following score equation using the one-step Newton--Raphson method: 
\begin{align*}
\mbf{0}
&=
\sum_{i=1}^n \sum_{s=1}^m
\mathds{1}(u_s\le \tau_{i,Q_i})\,\widetilde E(W_{ijks})\\
&\qquad \times
\left[
\begin{pmatrix}
\mbf{X}_i(u_s)\\
\hat{\mbf{\Xi}}_i
\end{pmatrix}
-
\frac{
\sum_{i'=1}^n \mathds{1}(u_s\le \tau_{i',Q_{i'}})
\begin{pmatrix}
\mbf{X}_{i'}(u_s)\\
\hat{\mbf{\Xi}}_{i'}
\end{pmatrix}
\exp\left\{
\mbf{\gamma}_{jk}^{\T}\mbf{X}_{i'}(u_s)
+\mbf{B}_{jk}^{\T}\hat{\mbf{\Xi}}_{i'}
\right\}
\widetilde E\{\exp(b_{i'})\}
}{
\sum_{i'=1}^n \mathds{1}(u_s\le \tau_{i',Q_{i'}})
\exp\left\{
\mbf{\gamma}_{jk}^{\T}\mbf{X}_{i'}(u_s)
+\mbf{B}_{jk}^{\T}\hat{\mbf{\Xi}}_{i'}
\right\}
\widetilde E\{\exp(b_{i'})\}
}
\right].
\end{align*}
Finally, $\sigma^2$ is updated by $(\sigma^2)^{\mathrm{new}}=n^{-1}\sum_{i=1}^n \widetilde E(b_i^2)$.

The E-step and M-step are iterated until convergence. 
Denote the resulting estimators for $\mbf{\theta}$, $\mathcal{B}_{r_n}$, and $\check{H}_{jk}$ by $\hat{\mbf{\theta}} = \left[ \{\hat{\mbf{\gamma}}_{jk}\}_{(j,k)\in\mathcal{D}}, \hat\sigma^2 \right]$, $\hat{\mathcal{B}}_{r_n}= \{\hat{\mbf{B}}_{jk}\}_{(j,k) \in \mathcal{D}}$, and $\bar{H}_{jk}$, respectively, where $\hat{\mbf{B}}_{jk} = (\hat{\beta}_{jk1}, \ldots, \hat{\beta}_{jkr_n})^{\T}$.
Then the original functional coefficients $\beta_{jk}(v)$ in model~\eqref{eq1} can be estimated by $\hat{\beta}_{jk}(v)=\sum_{l=1}^{r_n}\hat{\beta}_{jkl}\hat{\phi}_l(v)$, and the original cumulative baseline intensity function $H_{jk}$ can be estimated by $\hat{H}_{jk} = \bar{H}_{jk} \exp \left\{-\int_{\mathcal{V}}\hat{\beta}_{jk}(v)\hat{\mu}(v)dv\right\}$.

In practice, we select the optimal number of FPCs, $r_n$, based on the proportion of variance explained (PVE), defined as $\sum_{l=1}^{r_n}\lambda_l/\sum_{l=1}^{\infty}\lambda_l$. Although alternatives such as the Akaike Information Criterion (AIC) are available, our numerical studies suggest that AIC tends to select a larger $r_n$, resulting in increased estimation error and inflated Type I error.

\subsection{Profile score test}
\label{subsec:score_test}
In real applications, it is of great interest to test the overall association between the imaging covariates and the multistate process.
Traditional large-sample testing methods, such as the Wald and likelihood ratio tests, are not applicable to our semiparametric functional multistate model because the proposed estimators converge at rates slower than $n^{1/2}$ and are therefore not asymptotically normal; see Theorem~\ref{thm1: convergence_rate} in Section~\ref{sec:asymp_theory}.
To overcome this difficulty, we develop a score test based on the profile pseudo likelihood to test the global null hypothesis $H_0: \beta_{jk}(v) = 0$ for $v\in\mathcal{V}$ and $(j,k)\in\mathcal{D}$.
Consistent with the FPCA procedure in Section~\ref{subsec:fpca}, we truncate the infinite expansion of $\beta_{jk}(v)$ to the first $r_n$ terms and instead test $H_0^*: \mathcal{B}_{r_n} = \mbf{0}$.
To circumvent the challenge posed by the nuisance parameter $\mathcal{H}$, we borrow the idea of profile likelihood \citep{murphy2000profile} and consider the profile pseudo log-likelihood $\text{pl}_n(\mbf{\theta},\mathcal{B}_{r_n})=\max_{\mathcal{H}}\log \tilde L_n(\mbf{\theta},\mathcal{B}_{r_n},\mathcal{H})$, which can be computed using the same procedure as in Section~\ref{subsec:npmple}, with $\mbf{\theta}$ and $\mathcal{B}_{r_n}$ held fixed.
We construct our score test statistic using the score function and information matrix derived from $\text{pl}_n(\mbf{\theta},\mathcal{B}_{r_n})$.
Specifically, let $\tilde{\mbf{\theta}} = \arg\max_{\mbf{\theta}}\text{pl}_n(\mbf{\theta}, \mbf{0})$ denote the constrained nonparametric maximum pseudo likelihood estimator of $\mbf{\theta}$ under $H_0^*$.
We approximate the score and information at $(\tilde{\mbf{\theta}}, \mbf{0})$ through numerical differentiation of $\text{pl}_n(\mbf{\theta},\mathcal{B}_{r_n})$:
\[
\mbf{U}_n(\tilde{\mbf{\theta}}, \mbf{0}) = \nabla_{\delta_n}\text{pl}_n(\tilde{\mbf{\theta}}, \mbf{0}) \quad \text{ and } \quad \mathcal{I}_n(\tilde{\mbf{\theta}}, \mbf{0})=\sum_{i=1}^n\left\{\nabla_{\delta_n}\text{pl}_{ni}(\tilde{\mbf{\theta}}, \mbf{0})\right\}^{\otimes2},
\]
where $\nabla_{\delta_n}$ denotes the first-order numerical derivative with perturbation constant $\delta_n = O(n^{-1/2})$, $\text{pl}_{ni}(\tilde{\mbf{\theta}}, \mbf{0})$ denotes the $i$th subject's contribution to $\text{pl}_n(\tilde{\mbf{\theta}}, \mbf{0})$, and $\mbf{a}^{\otimes2} = \mbf{a}\mbf{a}^{\T}$ for a column vector $\mbf{a}$.
The score test statistic is 
\[
T_n = \mbf{U}_n(\tilde{\mbf{\theta}}, \mbf{0})^{\T} \mathcal{I}^{-1}_n(\tilde{\mbf{\theta}}, \mbf{0}) \mbf{U}_n(\tilde{\mbf{\theta}}, \mbf{0}).
\]
We will establish the asymptotic null distribution of $T_n$ later in Section~\ref{sec:asymp_theory}.
In practice, we approximate the null distribution of $T_n$ by a chi-squared distribution with $|\mathcal{D}|r_n$ degrees of freedom, where $|\mathcal{D}|$ denotes the cardinality of $\mathcal{D}$.


\section{Asymptotic theory}
\label{sec:asymp_theory}
We now study the asymptotic properties of the proposed estimators and profile score test statistic. We first introduce some notation.
Let $\mbf{\theta}_0$, $\beta_{jk0}$, and $H_{jk0}$ denote the true values of $\mbf{\theta}$, $\beta_{jk} $, and $H_{jk}$, respectively, for $(j,k)\in\mathcal{D}$.
Let $\|\cdot\|$ denote the Euclidean norm for vectors, $\|\beta\|_{L_2} = \bigl\{\int_{\mathcal{V}}\beta(v)^2dv\bigr\}^{1/2}$ denote the $L_2$ norm for a functional coefficient $\beta(v)$ defined on $\mathcal{V}$, and $\|H\|_{L_2} = \bigl\{\int_0^{\tau}H(t)^2dt\bigr\}^{1/2}$ denote the $L_2$ norm for a function $H$ defined on $[0,\tau]$, where $[0,\tau]$ is the union of the supports of the examination times $\tau_{iq}$ ($i=1,\dots,n$ and $q=0,\dots,Q_i$).

We establish the consistency and mixed convergence rates of the proposed estimators in the following theorem.
\begin{theorem} \label{thm1: convergence_rate}
    Under Conditions 1--10 in Section~\SuppSecConditions\ of the Supplementary Material, for any $\epsilon \in (0, a_2/2)$,
    \[
    \|\hat{\mbf{\theta}} - \mbf{\theta}_{0}\| + \sum_{(j,k)\in\mathcal{D}} \left(\|\hat{\beta}_{jk} - \beta_{jk0}\|_{L_2} + \|\hat{H}_{jk} - H_{jk0}\|_{L_2}\right) =O_p\left(r_n^{a_1/2 + 3/4 + \epsilon} n^{-1/4} + r_n^{-a_2/2 + \epsilon}\right),
    \]
    where $a_1$ and $a_2$ are positive constants introduced in Conditions 1 and 2, and the right-hand side is $o_p(1)$.
\end{theorem}

Theorem~\ref{thm1: convergence_rate} shows that the convergence rates of the proposed estimators are only slightly slower than $r_n^{a_1/2 + 3/4} n^{-1/4} + r_n^{-a_2/2}$.
The dominant source of estimation bias arises from the FPCA step for handling the functional covariates and coefficients, where we truncate the number of FPCs at $r_n$ and replace the true FPC scores with their estimates.
Although our estimation framework builds on semiparametric techniques for multistate models, the theoretical analysis requires new arguments because our estimators are based on a pseudo likelihood rather than the likelihood under the true underlying model.
This model discrepancy introduces additional approximation error and further complicates the already challenging asymptotic theory for semiparametric multistate models under interval censoring. For example, a major technical challenge is to quantify the FPCA-induced approximation error in the pseudo log-likelihood, which is expressed through a product-integral representation.
We address these difficulties through a novel synthesis of techniques from FPCA theory, empirical process theory, and product integration \citep{hall2006properties,hall2007methodology,vaart1996weak,andersen1993statistical}.
The proof of Theorem~\ref{thm1: convergence_rate} is provided in Section~\SuppSecProofThmOne\ of the Supplementary Material.

The next theorem states the asymptotic null distribution of the proposed profile score test statistic $T_n$.
\begin{theorem} \label{thm:asymp_dist_Tn}
    Under Conditions 1--10 in Section~\SuppSecConditions\ of the Supplementary Material, we have $(T_n - |\mathcal{D}|r_n)/(2|\mathcal{D}|r_n)^{1/2} \overset{d}{\to}  N(0,1)$ under $H_0$.
\end{theorem}

Although the proposed estimators do not attain the optimal $n^{1/2}$ convergence rates, our proposed profile score statistic is nevertheless asymptotically normal under $H_0$.
Similar properties have been established for score tests in parametric settings with FPCA-processed functional covariates, such as functional linear regression and functional Cox regression with right-censored data \citep{kong2016classical,kong2018flcrm}.
However, the theoretical development in our semiparametric testing framework is substantially more challenging, due to the presence of the infinite-dimensional nuisance parameters $\mathcal{H}$.
We address these difficulties through a novel use of semiparametric efficiency theory \citep{bickel1993efficient}. For example, we employ the least favorable direction and the associated efficient score and information to orthogonalize the parameters of interest from the nuisance components. We further prove that $ \mbf{U}_n(\tilde{\mbf{\theta}}, \mbf{0}) $ and $\mathcal{I}_n(\tilde{\mbf{\theta}}, \mbf{0})$ provide consistent estimators for the efficient score and information using the profile likelihood theory \citep{murphy2000profile}.
The proof of Theorem~\ref{thm:asymp_dist_Tn} is provided in Section~\SuppSecProofThmTwo\ of the Supplementary Material.

\section{Simulation studies}
\label{sec:numerical}

We consider a three-state model with feasible transitions $\mathcal{D} = \{(1,2), (2,3)\}$ that mimics AD progression.
The cumulative baseline transition intensities are specified as $H_{12}(t)=\log(1 + 0.3t)$ and $H_{23}(t)=0.3t$.
The covariate vector is $\mbf{X}_i=(X_{i1},X_{i2})^{\T}$, with $X_{i1}\sim \text{Bernoulli}(0.5)$ and $X_{i2}\sim\text{Unif}(0,1)$.
The corresponding scalar coefficients are $\mbf{\gamma}_{12}=(0.5,-0.5)^{\T}$ and $\mbf{\gamma}_{23}=(0.4,0.2)^{\T}$.
The functional covariate is generated as $M_i(v)=u_{i1}+u_{i2}v+\sum_{j = 1}^{10}[w_{ij1}\sin\{(2j - 1)\pi v\}+w_{ij2}\cos\{(2j - 1)\pi v\}]$, where $u_{i1},\, u_{i2}\sim N(0,1)$ and $w_{ij1},\, w_{ij2}\sim N(0,j^{-2})$.
The corresponding functional coefficients are specified as
$\beta_{12}(v)=C[\sin(\pi v)-\cos(\pi v)+\sin(3\pi v)-\cos(3\pi v)
+\sin(5\pi v)/9-\cos(5\pi v)/9+\sin(7\pi v)/16-\cos(7\pi v)/16
+\sin(9\pi v)/25-\cos(9\pi v)/25+(0.18\pi)^{-1/2}\exp\{-(v - 0.5)^2/0.18\}]$ and
$\beta_{23}(v)=C[\sin(\pi v)+\cos(\pi v)+\sin(3\pi v)+\cos(3\pi v)
+\sin(5\pi v)/9+\cos(5\pi v)/9+\sin(7\pi v)/16+\cos(7\pi v)/16
+\sin(9\pi v)/25+\cos(9\pi v)/25+(0.18\pi)^{-1/2}\exp\{-(v - 0.5)^2/0.18\}]$,
where the common multiplicative factor $C$ controls the overall effect size of the functional covariate and is set to $C = 0.3$ for estimation performance evaluation. Finally, we generate the random effect $b_i\sim N(0,\sigma^2)$ with $\sigma^2 = 0.64$. Each subject's initial state is randomly assigned as either 1 or 2 with equal probability. We then generate six potential examination times for each subject with the first being $\text{Unif}(0, 1)$, and the gap between any two successive examination times being $0.05 + \text{Unif}(0, 1)$. The study end time is set to $\tau = 3$, and any examination after $\tau$ is discarded.

\begin{figure}
    \centering
    \includegraphics[width=0.8\textwidth]{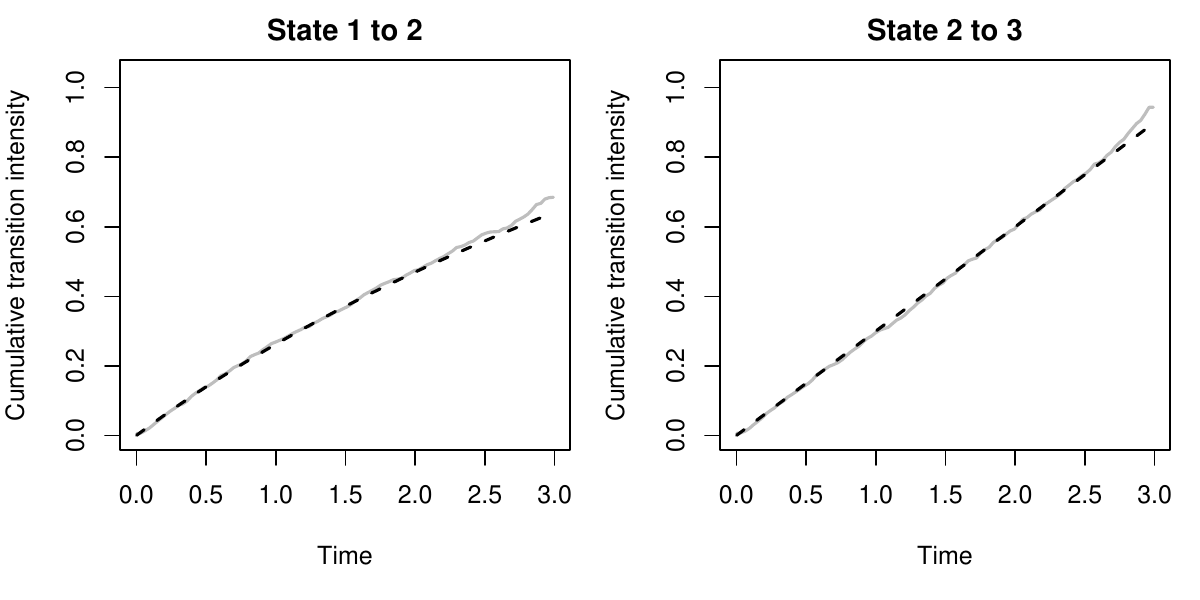}
    \includegraphics[width=0.8\textwidth]{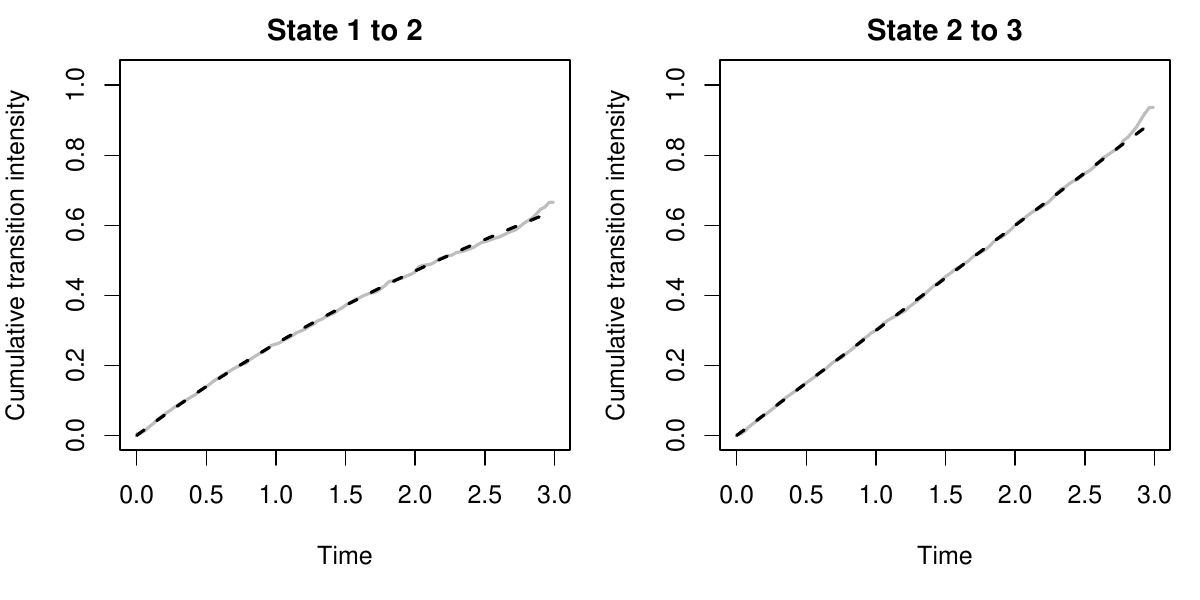}
    \caption{Estimation of the cumulative baseline transition intensities in the simulation studies. The solid gray line and dashed black curves respectively show the true values and the median estimates based on 500 replicates with sample sizes $n=400$ (top row) and $n=800$ (bottom row), where $r_n$ is selected such that at least 85\% of the total variance is explained.}
    \label{fig:Sim-cum-tran-intensity}
\end{figure}

We generate 500 replicates with sample sizes $n=400$ and $800$.
Table~\ref{tab:rmse_parameters} presents the relative mean squared errors for the scalar parameters $\mbf{\theta}$ and the functional parameters $\beta_{12}(v)$ and $\beta_{23}(v)$ across various choices for the truncation number $r_n$, including fixed $r_n\in\{1,\dots,8\}$ and the smallest $r_n$ such that PVE is at least 85\%. The estimation errors for the scalar parameters are quite stable across different choices of $r_n$, whereas the errors for the functional coefficients display a U-shaped pattern: they decrease up to $r_n= 4$ and increase thereafter. Selecting $r_n$ by PVE performs nearly as well as the optimal fixed choice $r_n=4$.
The estimated cumulative baseline transition intensities are presented in Figure~\ref{fig:Sim-cum-tran-intensity}. Notably, the estimated curves are almost identical to the ground truth.

We further evaluate the type I error and power of the proposed profile score test by varying the factor $C$, which regulates the overall functional effect in $\beta_{12}(v)$ and $\beta_{23}(v)$ from 0 to 0.1.
Table~\ref{tab:power} presents the empirical rejection rates at the 0.05 significance level.
For all fixed choices of $r_n$ and for the PVE-based selection, the type I error is well controlled.
As $C$ increases, the power increases rapidly for all choices of $r_n$.

\begin{table}
\caption{Relative mean squared errors of parameter estimates}
\label{tab:rmse_parameters}
\begin{center}
\begin{tabular}{ccccc}
\hline
 & $r_n$ & $\mbf{\theta}$ & $\beta_{12}(v)$ & $\beta_{23}(v)$ \\ \hline
$n=400$ & 1  & 0.240 (0.428)  & 0.216 (0.063)  & 0.373 (0.070) \\
 & 2  & 0.253 (0.476)  & 0.191 (0.106)  & 0.254 (0.126) \\
 & 3  & 0.267 (0.519)  & 0.174 (0.136)  & 0.156 (0.111) \\
 & 4  & 0.276 (0.612)  & 0.170 (0.159)  & 0.105 (0.113) \\
 & 5  & 0.292 (0.627)  & 0.314 (0.334)  & 0.193 (0.202) \\
 & 6  & 0.308 (0.722)  & 0.520 (0.510)  & 0.306 (0.293) \\
 & 7  & 0.313 (0.716)  & 0.890 (0.878)  & 0.530 (0.529) \\
 & 8  & 0.319 (0.724)  & 1.434 (1.232)  & 0.796 (0.818) \\
 & PVE & 0.276 (0.612) & 0.170 (0.159)  & 0.105 (0.113) \\ \hline
$n=800$ & 1  & 0.136 (0.119)  & 0.198 (0.040)  & 0.363 (0.046) \\
 & 2  & 0.149 (0.153)  & 0.162 (0.066)  & 0.216 (0.107) \\
 & 3  & 0.159 (0.180)  & 0.123 (0.067)  & 0.128 (0.090) \\
 & 4  & 0.170 (0.193)  & 0.095 (0.064)  & 0.068 (0.043) \\
 & 5  & 0.171 (0.196)  & 0.183 (0.150)  & 0.117 (0.091) \\
 & 6  & 0.172 (0.194)  & 0.284 (0.218)  & 0.176 (0.125) \\
 & 7  & 0.173 (0.190)  & 0.484 (0.364)  & 0.276 (0.187) \\
 & 8  & 0.174 (0.190)  & 0.743 (0.531)  & 0.430 (0.292) \\
 & PVE & 0.170 (0.193) & 0.095 (0.064)  & 0.068 (0.043) \\ \hline
\end{tabular}
\end{center}
\par\vspace{6mm}
\footnotesize\noindent
\parbox{0.98\linewidth}{PVE, selecting $r_n$ such that at least 85\% of the total variance is explained. The relative mean squared errors for $\mbf{\theta}$ and $\beta_{jk}(v)$ are defined as $\|\hat{\mbf{\theta}}-\mbf{\theta}_0\|^2 / \|\mbf{\theta}_0\|^2$ and $\|\hat{\beta}_{jk}-\beta_{jk0}\|_{L_{2}}^{2}/\|\beta_{jk0}\|_{L_{2}}^{2}$, respectively. Empirical standard errors of the relative squared errors are reported in parentheses. All numbers are based on 500 replicates.}
\end{table}

\begin{table}
\caption{Rejection rates of profile score test}
\label{tab:power}
\centering
\begin{tabular}{ccccccccc}
\hline
$n$ & $r_n$ & $C=0$ & $C=0.02$ & $C=0.04$ & $C=0.06$ & $C=0.08$ & $C=0.1$ \\ \hline
400 & 1 & 0.05 & 0.08 & 0.19 & 0.37 & 0.62 & 0.82 \\
 & 2 & 0.06 & 0.08 & 0.16 & 0.33 & 0.55 & 0.77 \\
 & 3 & 0.05 & 0.07 & 0.14 & 0.29 & 0.51 & 0.74 \\
 & 4 & 0.05 & 0.08 & 0.12 & 0.27 & 0.47 & 0.70 \\
 & 5 & 0.05 & 0.06 & 0.10 & 0.23 & 0.43 & 0.67 \\
 & 6 & 0.05 & 0.06 & 0.10 & 0.23 & 0.41 & 0.63 \\
 & 7 & 0.05 & 0.06 & 0.09 & 0.22 & 0.37 & 0.59 \\
 & 8 & 0.05 & 0.07 & 0.10 & 0.21 & 0.35 & 0.57 \\
 & PVE & 0.05 & 0.08 & 0.12 & 0.27 & 0.47 & 0.70 \\ \hline
800 & 1 & 0.07 & 0.14 & 0.37 & 0.68 & 0.92 & 1.00 \\
 & 2 & 0.05 & 0.09 & 0.33 & 0.63 & 0.91 & 1.00 \\
 & 3 & 0.05 & 0.11 & 0.28 & 0.56 & 0.87 & 1.00 \\
 & 4 & 0.06 & 0.12 & 0.26 & 0.54 & 0.85 & 1.00 \\
 & 5 & 0.06 & 0.10 & 0.26 & 0.50 & 0.80 & 1.00 \\
 & 6 & 0.06 & 0.09 & 0.22 & 0.48 & 0.77 & 1.00 \\
 & 7 & 0.05 & 0.07 & 0.21 & 0.45 & 0.73 & 1.00 \\
 & 8 & 0.06 & 0.08 & 0.19 & 0.44 & 0.70 & 1.00 \\
 & PVE & 0.06 & 0.12 & 0.26 & 0.54 & 0.85 & 1.00 \\ \hline
\end{tabular}
\par\vspace{8mm}
\footnotesize\noindent
\parbox{0.98\linewidth}{PVE, selecting $r_n$ such that at least 85\% of the total variance is explained. Reported rejection rates are based on 500 replicates.}
\end{table}

\section{Application}
\label{subsec:adni}
We now apply the proposed methods to the ADNI data to investigate the association between baseline lateral ventricular morphology and AD progression and to predict participants’ future disease trajectories. 

\subsection{ADNI dataset and preprocessing}
Our analysis uses ADNI participants with available baseline measurements within the lateral ventricle ROI, obtained through the preprocessing and quality-control pipeline described in Section~\ref{MRI_preprocessing}. This processed cohort comprises 768 participants with a mean follow-up of 3.1 years, including 218 CN, 375 MCI, and 175 AD participants at baseline. Because AD is treated as an absorbing state in the proposed progressive model, the 175 participants who already have AD at baseline provide no information about the model parameters and are therefore excluded from the analysis. Among the remaining 593 baseline-CN or baseline-MCI participants, 38 participants exhibit at least one reverse transition in their observed diagnosis histories and are excluded. The resulting analysis cohort contains 555 participants, including 205 baseline-CN and 350 baseline-MCI participants, with a mean follow-up of 5.4 years. Within this cohort, 50 CN-to-MCI changes, 217 MCI-to-AD changes, and 3 CN-to-AD changes are observed across successive visits. 
Baseline scalar covariates include age (years), gender (male versus female), years of education, marital status (married versus others), and APOE-$\varepsilon$4 carrier status (at least one allele versus no allele). 

\subsection{MRI Data Preprocessing}
\label{MRI_preprocessing}
Raw structural MRI data are acquired from the ADNI database (https://adni.loni.usc.edu/). To prepare the data for subsequent voxel-wise statistical analysis, we implement a standardized preprocessing pipeline relying on Advanced Normalization Tools \citep{avants2011reproducible}.
The specific preprocessing steps are conducted as follows:
\begin{enumerate}
    \item Bias Correction and Brain Extraction: We first apply N4 bias field correction to address intensity non-uniformities. This is followed by registration-based brain extraction (skull stripping) to isolate brain tissue.
    \item Tissue Segmentation: We utilize a prior-based N4-Atropos segmentation approach with the OASIS template to classify the brain into six distinct compartments: white matter (WM), gray matter (GM), deep GM, cerebrospinal fluid (CSF), brainstem, and cerebellum.
    \item Registration and Parcellation: Images are spatially registered to the uniform-resolution OASIS30-Atropos template. Next, we perform multi-atlas cortical parcellation on the skull-stripped and N4-corrected images, utilizing the 101 manually edited ROIs provided by the public MindBoggle-101 dataset \citep{klein2012101}.
    \item Quality Control: All processed images undergo manual quality control. Subjects demonstrating inaccurate registration or segmentation are excluded from the study. Furthermore, three specific ROIs are removed from the analysis due to a high frequency of missing values.
    \item Transformation Map Extraction: For the remaining subjects, we extract the Log-Jacobian transformation map in standard space. These maps are parcellated into the final 98 ROIs (the original 101 minus the 3 excluded ones) and cropped to a uniform dimension of $133 \times 129 \times 179$.
    \item Functional Covariate Construction: We focus on the lateral ventricle ROI. The 408 voxels within this ROI are extracted from the Log-Jacobian transformation map and ordered according to their spatial locations to construct the one-dimensional functional imaging covariate used in our model.
\end{enumerate}
For a more comprehensive discussion of this specific methodology, we refer readers to \cite{zhao2019genome}.
Following the above preprocessing pipeline, baseline lateral ventricle imaging data are eventually represented as a one-dimensional functional covariate with 408 voxels.

\subsection{Testing and model fitting}
We first test the global null hypothesis that lateral ventricular morphology has no effect on either transition intensity. The profile score test is highly significant ($p<0.005$) across all truncation numbers $r_n=1,\dots,5$, suggesting that baseline lateral ventricular morphology carries substantial prognostic information about subsequent disease progression.

FPCA of these spatially ordered voxels within the lateral ventricle ROI reveals that the first two components explain 95\% of the total variation and the first four components capture 99\%. We focus on presenting results based on $r_n=4$, while results for $r_n=2$ are consistent.
Table~\ref{tab:ADNI_PVE_4} presents the estimated effects of the baseline scalar covariates. As expected, APOE-$\varepsilon$4 carriage strongly accelerates progression in both transitions. Conditional on the other covariates and the shared random effect, carriers have estimated transition intensities approximately 2.12 times as high for CN-to-MCI progression and 2.25 times as high for MCI-to-AD progression as noncarriers. Older age is associated with faster CN-to-MCI progression ($p=0.038$) but shows no association with MCI-to-AD progression. Gender, years of education, and marital status show no clear associations with either transition.
The variance of the random effect is estimated as $0.261$ with an estimated standard error of $0.212$, indicating moderate dependence between the transition from CN to MCI and the transition from MCI to AD. The relatively large standard error is expected as only 17 participants are observed to experience both transitions by the end of follow-up.

\begin{table}[!t]
\centering
\small
\caption{Estimated scalar covariate effects in the ADNI study.}
\label{tab:ADNI_PVE_4}
\setlength{\tabcolsep}{5pt}
\begin{tabular}{lcccccc}
\toprule
& \multicolumn{3}{c}{CN to MCI}
& \multicolumn{3}{c}{MCI to AD} \\
\cmidrule(lr){2-4}\cmidrule(lr){5-7}
Covariate
& Estimate & SE & \(p\)-value
& Estimate & SE & \(p\)-value \\
\midrule
Age
&  0.022 & 0.010 & 0.038
&  0.000 & 0.005 & 0.992 \\
Gender
&  0.407 & 0.342 & 0.235
& -0.271 & 0.174 & 0.121 \\
Education years
& -0.049 & 0.052 & 0.349
&  0.019 & 0.023 & 0.402 \\
Married
& -0.429 & 0.337 & 0.203
&  0.219 & 0.205 & 0.286 \\
APOE-$\varepsilon4$ carrier
&  0.750 & 0.313 & 0.017
&  0.812 & 0.164 & \(<0.001\) \\
\bottomrule
\end{tabular}
\end{table}

\begin{figure}
    \centering
    \includegraphics[width=0.9\linewidth]{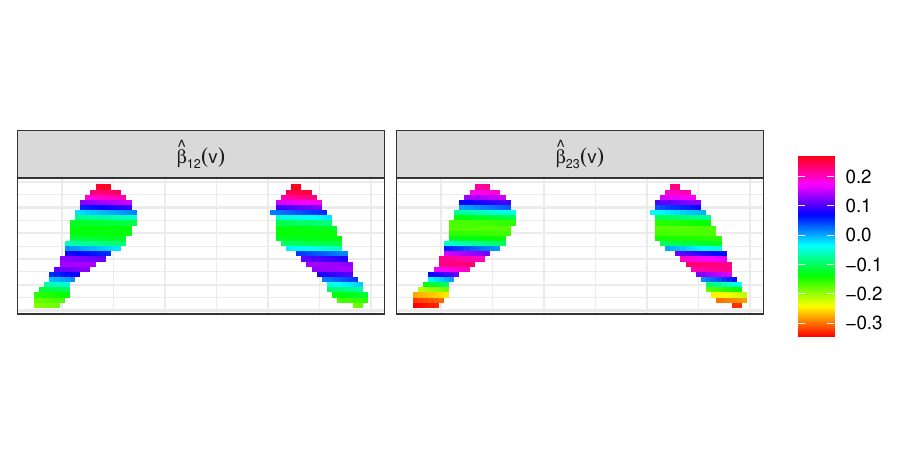}
    \vspace{-2cm}
    \caption{Estimated functional coefficients in the ADNI study.}
    \label{fig:adni_beta}
\end{figure}

To better interpret the imaging effects, Figure~\ref{fig:adni_beta} visualizes the estimated functional coefficients $\hat{\beta}_{12}(v)$ and $\hat{\beta}_{23}(v)$. Both panels show spatially heterogeneous associations between baseline ventricular morphology and subsequent disease progression, with broadly similar patterns across the left and right ventricles. 
Greater local expansion in the upper region (anterior horn) and a localized portion of the ventricular body is associated with higher transition intensity for both transitions. Because ventricular enlargement is an indirect marker of surrounding cerebral tissue loss, this pattern may reflect regionally distributed neurodegeneration rather than uniform global ventricular enlargement. Similar abnormalities in the anterior horn and body/occipital regions have been reported in previous ventricular morphometry studies \citep{apostolova2012hippocampal,dong2020applying}.
The positive association in the middle region is stronger for the MCI-to-AD transition. This may indicate that, once a participant has reached the MCI stage, ventricular-body morphology provides additional information about the more widespread cerebral atrophy associated with conversion to AD. Previous studies have similarly reported greater frontal and body enlargement among MCI converters and more advanced AD cases \citep{apostolova2012hippocampal,apostolova2013ventricular}.
In contrast, the estimated coefficients in the lower region (inferior horn) are negative, particularly for the MCI-to-AD transition. This distinct pattern may partly arise because this hippocampus-adjacent area is affected relatively early in the disease process and may provide less additional information at baseline when more widespread ventricular enlargement is already present \citep{jack2010hypothetical}.
Nevertheless, because ventricular morphology is represented jointly through an FPC basis and the anatomical regions are strongly correlated, this pattern should be interpreted as a relative morphological contrast between the middle and lower ventricular regions, rather than as evidence that inferior-horn enlargement is protective.

For transition $(j,k)$, we summarize the functional imaging contribution by the imaging risk score, defined as $\int \hat{\beta}_{jk}(v)
\{M_i(v)-\hat{\mu}(v)\}\,dv
=
\sum_{l=1}^{4}
\hat{\beta}_{jkl}\hat{\xi}_{il}$.
Figure~\SuppFigCumHaz\ in the Supplementary Material presents the estimated cumulative transition intensity functions stratified by APOE-$\varepsilon4$ carrier status and by selected percentiles of the imaging risk score. The clear separation of the cumulative transition intensities across these groups illustrates the heterogeneity in disease progression captured by the fitted model. For the MCI-to-AD transition, the participants corresponding to the 25th, 50th, and 75th percentiles of the imaging risk score have ventricular volumes of 44,700, 58,718, and 47,760~mm$^3$, respectively. This non-monotone relationship suggests that the imaging risk score captures spatial patterns of ventricular morphology beyond overall ventricular volume. 
Figure~\SuppFigTransProb\ in the Supplementary Material displays the estimated five-year transition probabilities from CN and MCI to different states under the fitted model, with the covariates equal to the sample medians. As expected, the probabilities of progression toward more severe states generally increase over time.

\subsection{Model comparison in prediction}
We focus on MCI-to-AD prediction because existing approaches
primarily target this transition. In addition, the number of CN-to-MCI transitions is relatively small, which may lead to unstable evaluation in validation sets.
The proposed Functional multistate model (MSM) with four FPC scores is compared with three alternatives: (i) a No-imaging MSM that excludes imaging information; (ii) a Volume MSM that replaces the functional imaging data with a scalar ventricular-volume summary; and (iii) a Functional Cox model that includes the same four FPC scores as Functional MSM and is fitted for MCI-to-AD progression using baseline-MCI participants only. All four models include the same scalar covariates. 
The Functional Cox model was originally developed for right-censored data \citep{kong2018flcrm}. However, because the ADNI transition times are interval-censored and the prediction metrics considered below are defined for interval-censored outcomes, we extend this approach to interval-censored data using the estimation procedure of \citet{zeng2016maximum}, as implemented in the R package \texttt{IntCens}. This provides a fair comparison between the three multistate models and Functional Cox under the same observation scheme and evaluation criteria.

We divide the 350 baseline-MCI participants into training and validation sets at a 9:1 ratio over 200 random splits. Splitting is stratified by transition status
to preserve the transition-to-censoring ratio across replicates and reduce variability caused by the small validation sample size. In addition, all 205 baseline-CN participants are retained in every training set.

To evaluate predictive performance, we consider two metrics: negative log-likelihood (NLL) and integrated Brier score (IBS). Let \(\hat S_{im}(t)\) denote the estimated probability that subject \(i\) remains free of the MCI-to-AD transition through time \(t\) under model \(m\). For the three multistate models, this probability is marginalized over the estimated shared random-effect distribution. 
Then for subject $i$ with transition time \(T_i\in(L_i,R_i]\), the estimated probability of the observed outcome is $\hat p_{im}
=
\hat S_{im}(L_i)-\hat S_{im}(R_i)\mathds{1}(R_i<\infty)$.
For a validation set \(\mathcal V\) of size \(n_{v}\), the NLL is defined as
\[
\operatorname{NLL}_{m}
=
-\frac{1}{n_v}
\sum_{i\in\mathcal V}
\log\left\{\max(\hat p_{im},10^{-15})\right\}.
\]

To calculate the IBS for interval-censored outcomes \(T_i\in(L_i,R_i]\), we first define a pseudo-survival status as
\[
q_i(t)
=
\mathds{1}(t\le L_i)
+
\mathds{1}(L_i<t\le R_i)
\frac{
\hat S_{\mathrm{ref}}(t)-\hat S_{\mathrm{ref}}(R_i)
}{
\hat S_{\mathrm{ref}}(L_i)-\hat S_{\mathrm{ref}}(R_i)
},
\]
where \(\hat S_{\mathrm{ref}}(t)\) denotes the Turnbull survival estimator \citep{turnbull1976empirical} obtained from the full baseline-MCI cohort. We fix \(\hat S_{\mathrm{ref}}\) across models and replicates and set \(\hat S_{\mathrm{ref}}(\infty)=0\).
Then for model \(m\), the IBS up to time horizon \(\tau\) is defined as
\[
\operatorname{IBS}_m
=
\frac{1}{n_v\,\tau}\sum_{i\in\mathcal V}
\int_0^{\tau}
\left\{q_i(t)-\hat S_{im}(t)\right\}^2\,dt.
\]
We report IBS over the entire observed MCI-to-AD follow-up period (\(\tau=18.53\) years). Smaller NLL and IBS values indicate better predictive performance.

\begin{table}[!t]
\centering
\small
\setlength{\tabcolsep}{6pt}
\caption{Prediction performance for MCI-to-AD progression.}
\label{tab:prediction-performance}
\begin{tabular}{lcccc}
\toprule
Metric & Functional MSM 
& No-imaging MSM
& Volume MSM
& Functional Cox \\
\midrule
NLL
& \textbf{2.154 (0.657)}
& 2.193 (0.676)
& 2.163 (0.667)
& 2.365 (0.822) \\
IBS
& \textbf{0.110 (0.019)}
& 0.111 (0.017)
& \textbf{0.110 (0.017)}
& 0.111 (0.018) \\
\bottomrule
\end{tabular}
\begin{flushleft}
\footnotesize
Values are means across 200 replicates, with standard deviations across replicates in parentheses.
\end{flushleft}
\end{table}

Table~\ref{tab:prediction-performance} presents the prediction results for all four models based on 200 replicates. The proposed Functional MSM achieves the best overall predictive performance, with the lowest NLL and competitive IBS among all competing methods. These results suggest that jointly modeling the multistate disease process with functional imaging information can improve prediction of MCI-to-AD progression.

\section{Discussion}
\label{s:discuss}

In this paper, we propose a semiparametric functional multistate model that enables prediction of future AD progression from any current disease state using imaging biomarkers. We establish asymptotic properties of the proposed estimators, demonstrate satisfactory finite-sample performance through simulation studies, and identify transition-specific associations between baseline lateral ventricular morphology and AD progression in the ADNI application. By modeling the entire disease course and using all available transition information, our model achieves the best overall predictive performance among the competing methods.
Although motivated by Alzheimer's disease, the proposed modeling framework is broadly applicable to other chronic diseases with rich imaging or functional biomarkers, such as Parkinson's disease, multiple sclerosis, and retinal degenerative diseases.


The ADNI analysis illustrates that the proposed framework can extract interpretable stage-specific imaging effects from interval-censored multistate data. One limitation, however, is that death is a potential semi-competing terminal event, which is not modeled in the current analysis. In our exploratory checks, recorded death proportions are broadly similar across the CN, MCI, and AD groups, providing some support to the noninformative-censoring assumption used here. Moreover, among subjects with available cause-of-death information, only very few deaths are explicitly attributed to Alzheimer’s disease, while most deaths are recorded as due to other causes. Thus, we view semi-competing death as a limitation of the ADNI application, but not one that appears to materially alter the main findings. A full treatment of death would require a substantially more complex extension of the current framework and is left for future work.

Another limitation is that model~\eqref{eq1} accommodates only imaging covariates measured at baseline or a fixed time point. An important direction for future work is to develop functional joint models that accommodate longitudinal imaging biomarkers, which would provide a more comprehensive characterization of disease dynamics and allow for fully dynamic prediction of future disease progression. Moreover, Alzheimer's disease progression is highly heterogeneous. It is therefore clinically meaningful to investigate latent subgroups among participants and to develop personalized prediction methods, for example via latent-class or mixed-membership extensions of our current framework.

\begin{funding}
Kang is partially supported by the National Natural Science Foundation of China (Grant No.\ 12301368).
Gu is partially supported by the Hong Kong Research Grant Council grant 27303624.
\end{funding}

\begin{supplement}
\stitle{Supplementary Material for ``Semiparametric Functional Multistate Modeling of Alzheimer’s Disease Progression with Imaging Biomarkers''}\sdescription{Supplementary Material includes technical details referenced in Section~\ref{subsec:npmple}, technical proofs of Theorems~\ref{thm1: convergence_rate} and \ref{thm:asymp_dist_Tn}, and additional details on the ADNI data application.}
\end{supplement}


\bibliographystyle{imsart-nameyear} 
\bibliography{paper-ref}       


\end{document}


\renewcommand{\thetheorem}{S.\arabic{theorem}}
\renewcommand{\thelemma}{S\arabic{lemma}}
\setcounter{theorem}{0}
\setcounter{lemma}{0}
\setcounter{condition}{0}

\begin{frontmatter}
\title{Supplementary Material for ``Semiparametric Functional Multistate Modeling of Alzheimer’s Disease Progression with Imaging Biomarkers''}

\begin{aug}
\author[A]{\fnms{Chenrui}~\snm{Qi}\ead[label=e1]{qichenrui@connect.hku.hk}}
\author[B]{\fnms{Kai}~\snm{Kang}\ead[label=e2]{kangk5@mail.sysu.edu.cn}}
\author[A]{\fnms{Yu}~\snm{Gu}\ead[label=e3]{yugu@hku.hk}}

\address[A]{Department of Statistics and Actuarial Science, University of Hong Kong, Pokfulam Road, Hong Kong\printead[presep={,\ }]{e1,e3}}
\address[B]{Department of Statistics, Sun Yat-sen University, Guangzhou, China\printead[presep={,\ }]{e2}}
\end{aug}

\end{frontmatter}

\section{Details of Estimation}
\label{S0:EM}
\subsection{Equivalence between the original and the Poisson-based pseudo likelihood}
\label{S0.1:Equivalence}
We show that the Poisson-based pseudo likelihood induced by the latent variables $\{W_{ijks}: i=1,\ldots,n,\ (j,k)\in\mathcal{D},\ s=1,\ldots,m\}$ 
is equivalent to the original pseudo likelihood. For notational convenience, denote
$
\omega_{ijks}
=
\Delta \check H_{jk}(u_s)
\exp\left\{
\mbf{\gamma}_{jk}^{\T}\mbf{X}_i(u_s)
+\mbf{B}_{jk}^{\T}\hat{\mbf{\Xi}}_i
+b_i
\right\},
(j,k)\in\mathcal{D},\ s=1,\ldots,m
$.
Consider one observed interval $(\tau_{i,q-1},\tau_{iq}]$ for subject $i$, where the observed states at the two endpoints are $S_{i,q-1}$ and $S_{iq}$, respectively.
Let $\tau_{i,q-1}=u_{s_0}<u_{s_1}<\cdots<u_{s_r}<u_{s_{r+1}}=\tau_{iq}$ 
be the ordered distinct examination times in the closed interval $[\tau_{i,q-1},\tau_{iq}]$. 
A feasible transition path from $S_{i,q-1}$ to $S_{iq}$ over $(\tau_{i,q-1},\tau_{iq}]$ is a sequence $(k_0,k_1,\ldots,k_r,k_{r+1})$, 
where $k_0=S_{i,q-1}$, $k_{r+1}=S_{iq}$, and for each $l=1,\ldots,r+1$, either $k_{l-1}=k_l$ or $(k_{l-1},k_l)\in\mathcal{D}$. Let $\mathcal{A}_{iq}$ denote the set of all such feasible $(k_1,\ldots,k_r)$. 

For a general continuous-time multistate process, consider a partition $\tau_{i,q-1}=t_0<t_1<\cdots<t_r<\tau_{iq}=t_{r+1}$ such that there is at most one transition within each $(t_{l-1},t_l], l=1,\ldots,r+1$. Let $j_0,\ldots, j_{r+1}$ denote the states occupied at $t_0,\ldots,t_{r+1}$ and let $\mathcal{A}_r$ denote the set of all feasible intermediate state sequences $(j_1,\ldots,j_r)$ connecting $S_{i,q-1}$ and $S_{iq}$. Then, the transition probability can be written as
\begin{align*}
&\mbf{P}_i(\tau_{i,q-1},\tau_{iq};\mbf{X}_i,\hat{\mbf{\Xi}}_i,b_i)^{(S_{i,q-1},S_{iq})} \\
&=
\sum_{(j_1,\ldots,j_r)\in\mathcal{A}_r}
\prod_{l=1}^{r+1}
\Bigg(
\exp\Bigg\{
-\sum_{k:(j_{l-1},k)\in\mathcal{D}}
\int_{t_{l-1}}^{t_l}
d{\mbf A}_i(t;\mbf{X}_i,\hat{\mbf{\Xi}}_i,b_i)^{(j_{l-1},k)}
\Bigg\}
\Bigg)^{\mathds{1}(j_{l-1}=j_l)}
\\
&\qquad\qquad\times
\Bigg[
\Bigg\{
1-\exp\Bigg(
-\int_{t_{l-1}}^{t_l}
d{\mbf A}_i(t;\mbf{X}_i,\hat{\mbf{\Xi}}_i,b_i)^{(j_{l-1},j_l)}
\Bigg)
\Bigg\}
\\
&\qquad\qquad\qquad\qquad\times
\exp\Bigg\{
-\sum_{\substack{k:(j_{l-1},k)\in\mathcal{D}\\ k\neq j_l}}
\int_{t_{l-1}}^{t_l}
d{\mbf A}_i(t;\mbf{X}_i,\hat{\mbf{\Xi}}_i,b_i)^{(j_{l-1},k)}
\Bigg\}
\Bigg]^{\mathds{1}(j_{l-1}\neq j_l)} .
\end{align*}
Under the working model and the nonparametric maximum likelihood estimation approach, we can replace $\{(t_0,\ldots,t_{r+1}), (j_0,\ldots,j_{r+1})\}$ by $\{ (u_{s_0}, \ldots,u_{s_{r+1}}), (k_0,\ldots,k_{r+1}) \}$ and plug in the discretized cumulative transition intensity functions to obtain 
\begin{align*}
&\mbf{P}_i(\tau_{i,q-1},\tau_{iq};\mbf{X}_i,\hat{\mbf{\Xi}}_i,b_i)^{(S_{i,q-1},S_{iq})} \\
&=
\sum_{(k_1,\ldots,k_r)\in\mathcal{A}_{iq}}
\prod_{l=1}^{r+1}
\left[
\exp\left\{
-\sum_{k':(k_{l-1},k')\in\mathcal{D}}
\omega_{ik_{l-1}k's_l}
\right\}
\right]^{\mathds{1}(k_{l-1}=k_l)}
\\
&\qquad\qquad\times
\left[
\left\{
1-\exp\left(-\omega_{ik_{l-1}k_ls_l}\right)
\right\}
\exp\left\{
-\sum_{\substack{k':(k_{l-1},k')\in\mathcal{D}\\ k'\neq k_l}}
\omega_{ik_{l-1}k's_l}
\right\}
\right]^{\mathds{1}(k_{l-1}\neq k_l)}.
\end{align*}

Next, we consider the Poisson path event. For a given feasible path $(k_0,\ldots,k_{r+1})$, we can define the event of latent trajectory  $V_i(k_1,\ldots,k_r;\tau_{i,q-1},\tau_{iq},S_{i,q-1},S_{iq})$ as follows: (i) for each $l=1,\ldots,r+1$, if $k_{l-1}\neq k_l$, then $W_{ik_{l-1}k_ls_l}>0$ and $ W_{ik_{l-1}k's_l}=0$ for all $ k'\neq k_l$ such that  $(k_{l-1},k')\in\mathcal{D};$
(ii) if $k_{l-1}=k_l$, then $ W_{ik_{l-1}k's_l}=0$ for all  $k'$ such that $(k_{l-1},k')\in\mathcal{D}$. 
That is, given a path, all relevant Poisson variables are zero except that
$W_{ik_{l-1}k_ls_l}$ is positive whenever a transition from state $k_{l-1}$ to state $k_l$ occurs at time $u_{s_l}$.
By independence of the Poisson variables conditional on $(\mbf{X}_i,\hat{\mbf{\Xi}}_i,b_i)$, for any feasible path
$(k_0,\ldots,k_{r+1})$,
\begin{align*}
&\Pr\left\{
V_i(k_1,\ldots,k_r;\tau_{i,q-1},\tau_{iq},S_{i,q-1},S_{iq})
\mid \mbf{X}_i,\hat{\mbf{\Xi}}_i,b_i
\right\} \\
&=
\prod_{l=1}^{r+1}
\left[
\exp\left\{
-\sum_{k':(k_{l-1},k')\in\mathcal{D}}
\omega_{ik_{l-1}k's_l}
\right\}
\right]^{\mathds{1}(k_{l-1}=k_l)}
\\
&\qquad\qquad\times
\left[
\left\{
1-\exp\left(-\omega_{ik_{l-1}k_l s_l}\right)
\right\}
\exp\left\{
-\sum_{\substack{k':(k_{l-1},k')\in\mathcal{D}\\ k'\neq k_l}}
\omega_{ik_{l-1}k's_l}
\right\}
\right]^{\mathds{1}(k_{l-1}\neq k_l)}.
\end{align*}
We then define the event $
Y_i(\tau_{i,q-1},\tau_{iq},S_{i,q-1},S_{iq})
=
\bigcup_{(k_1,\ldots,k_r)\in\mathcal{A}_{iq}}
V_i(k_1,\ldots,k_r;\tau_{i,q-1},\tau_{iq},S_{i,q-1},S_{iq})$. Because the events $\left\{
V_i(k_1,\ldots,k_r;\tau_{i,q-1},\tau_{iq},S_{i,q-1},S_{iq})
:
(k_1,\ldots,k_r)\in\mathcal{A}_{iq}
\right\}$ are mutually exclusive, we obtain
\[
\Pr\left\{
Y_i(\tau_{i,q-1},\tau_{iq},S_{i,q-1},S_{iq})
\mid \mbf{X}_i,\hat{\mbf{\Xi}}_i,b_i
\right\}
=
\mbf{P}_i(\tau_{i,q-1},\tau_{iq};\mbf{X}_i,\hat{\mbf{\Xi}}_i,b_i)^{(S_{i,q-1},S_{iq})}.
\]
Now define $O_i = \bigcap_{q=1}^{Q_i}
Y_i(\tau_{i,q-1},\tau_{iq},S_{i,q-1},S_{iq})$, 
which represents the entire observed transition history of subject $i$.
Then
\[
\Pr(O_i\mid \mbf{X}_i,\hat{\mbf{\Xi}}_i,b_i)
=
\prod_{q=1}^{Q_i}
\mbf{P}_i(\tau_{i,q-1},\tau_{iq};\mbf{X}_i,\hat{\mbf{\Xi}}_i,b_i)^{(S_{i,q-1},S_{iq})}.
\]
Therefore, the Poisson-based pseudo likelihood is equivalent to the original pseudo likelihood, and maximizing the original pseudo likelihood is equivalent to maximizing the Poisson-based pseudo likelihood.

\subsection{Conditional expectation of $W_{ijks}$ given $O_i$ and $b_i$}
\label{S0.2:conditional_expectation}
Fix subject $i$, transition pair $(j,k)\in\mathcal{D}$, and time point $u_s$ with $u_s\le \tau_{i,Q_i}$.
Suppose that $u_s\in(\tau_{i,q-1},\tau_{iq}]$ for some $q\in\{1,\ldots,Q_i\}$.
Recall that the unique time points in $[\tau_{i,q-1},\tau_{iq}]$ are $\tau_{i,q-1}=u_{s_0}<u_{s_1}<\cdots<u_{s_l}=u_s<\cdots<u_{s_{r}}<u_{s_{r+1}}=\tau_{iq}$ and the corresponding feasible latent states are $(S_{i,q-1}=k_0, k_1,\ldots,k_l,\ldots,k_r,k_{r+1}=S_{iq})$ with $(k_1,\ldots,k_l,\ldots,k_r) \in \mathcal{A}_{iq}$.
Since $W_{ijks}$ only pertains to the interval $(\tau_{i,q-1},\tau_{iq}]$, the conditional expectation $E(W_{ijks}\mid O_i,b_i)$ can be written as
\begin{align*}
    &\qquad E\left\{W_{ijks}\mid Y_i(\tau_{i,q-1},\tau_{iq}, S_{i,q-1},S_{iq}),b_i\right\} \\&= \sum_{w=1}^{\infty} \frac{w\times \Pr\{ (W_{ijks}=w ) \cap Y_i(\tau_{i,q-1},\tau_{iq}, S_{i,q-1},S_{iq}) \}}{\Pr(Y_i(\tau_{i,q-1},\tau_{iq}, S_{i,q-1},S_{iq}))} 
    \\
    &= \sum_{w=1}^{\infty} \sum_{(k_1,\ldots,k_r)\in \mathcal{A}_{iq}} \frac{w\times \Pr\{ (W_{ijks}=w ) \cap V_i(k_1,\ldots,k_r; \tau_{i,q-1}, \tau_{iq},S_{i,q-1},S_{iq}) \}}{ \mbf{P}_i(\tau_{i,q-1},\tau_{iq};\mbf{X}_i,\hat{\mbf{\Xi}}_i,b_i)^{(S_{i,q-1},S_{iq})} }
\end{align*}

For notational simplicity, throughout this subsection we suppress the conditioning on
$\mbf{X}_i,\hat{\mbf{\Xi}}_i,b_i$ in all transition probabilities.
Then, conditioning on whether the subject occupies state $j$ immediately before time $u_s$, we obtain
\begin{align*}
E(W_{ijks}\mid O_i,b_i)
&= \sum_{j'\neq j}
\frac{
\mbf{P}_i(\tau_{i,q-1},u_{s-1})^{(S_{i,q-1},j')}
\mbf{P}_i(u_{s-1},\tau_{iq})^{(j',S_{iq})}
}{
\mbf{P}_i(\tau_{i,q-1},\tau_{iq})^{(S_{i,q-1},S_{iq})}
} \left\{ \sum_{w=1}^{\infty} w \Pr(W_{ijks}=w) \right\} \\
&\quad+
\frac{
\mbf{P}_i(\tau_{i,q-1},u_{s-1})^{(S_{i,q-1},j)}
\mbf{P}_i(u_s,\tau_{iq})^{(k,S_{iq})}
}{
\mbf{P}_i(\tau_{i,q-1},\tau_{iq})^{(S_{i,q-1},S_{iq})}
}
\left\{ \sum_{w=1}^{\infty}w\Pr(W_{ijks}=w, W_{ijk's}=0, k'\neq j,k) \right\}
\\
&=\sum_{j'\neq j}
\frac{
\mbf{P}_i(\tau_{i,q-1},u_{s-1})^{(S_{i,q-1},j')}
\mbf{P}_i(u_{s-1},\tau_{iq})^{(j',S_{iq})}
}{
\mbf{P}_i(\tau_{i,q-1},\tau_{iq})^{(S_{i,q-1},S_{iq})}
}
\omega_{ijks}
\\
&\quad+
\frac{
\mbf{P}_i(\tau_{i,q-1},u_{s-1})^{(S_{i,q-1},j)}
\mbf{P}_i(u_s,\tau_{iq})^{(k,S_{iq})}
}{
\mbf{P}_i(\tau_{i,q-1},\tau_{iq})^{(S_{i,q-1},S_{iq})}
}
\omega_{ijks}
\exp\left\{
-\sum_{\substack{k':(j,k')\in\mathcal{D}\\k'\neq k}}
\omega_{ijk's}
\right\}.
\end{align*}
The first term corresponds to paths for which the subject is not in state $j$ immediately before $u_s$. In this case, $W_{ijks}$ is not constrained by the observed path event and contributes its Poisson mean $\omega_{ijks}$.
The second term corresponds to paths for which the subject is in state $j$ immediately before $u_s$ and makes a transition from $j$ to $k$ at $u_s$, which requires $W_{ijks}>0$ and all competing transition counts out of state $j$ at time $u_s$ to be zero.

Finally, the conditional expectation used in the E-step is
\[
\widetilde E(W_{ijks})
=
\int_{b_i} E(W_{ijks}\mid O_i,b_i)\,f(b_i\mid O_i)\,db_i,
\]
where the posterior density of $b_i$ given $O_i$ satisfies
\[
f(b_i\mid O_i)
\propto
\prod_{q=1}^{Q_i}
\mbf{P}_i(\tau_{i,q-1},\tau_{iq})^{(S_{i,q-1},S_{iq})}
f(b_i;\sigma^2).
\]
The above one-dimensional integral is approximated numerically by Gaussian--Hermite quadratures in the implementation of the EM algorithm.

\section{Theoretical Proofs} \label{S1:Theory}

\subsection{Conditions}
\label{S1.1:Conditions}
For notational simplicity, we consider a generic subject and omit the subscript $i$ for all random quantities.
With a slight abuse of notation, we use $C$ to denote a generic constant whose value may change from place to place. Throughout this supporting information, we use norm notation aligned with the main paper: $\|\cdot\|$ denotes the Euclidean norm for finite-dimensional vectors, $\|\cdot\|_{L_2}$ denotes the $L_2$ norm for square-integrable functions on either the imaging domain $\mathcal{V}$ or time domain $[0, \tau]$, and we extend these norms element-wise to collections of parameters and functions, with $\|\mathcal{B}_{r_n}\|=\big(\sum_{(j,k) \in \mathcal{D}} \|\mbf{B}_{jk}\|^2 \big)^{1/2}$ and $\|\mathcal{H}\|_{L_2}=\big(\sum_{(j,k)\in \mathcal{D}} \|H_{jk}\|^2_{L_2} \big)^{1/2}$.

The first set of conditions mainly concerns the functional covariate $M(v)$.
Without loss of generality, we assume that $M(v)$  is a mean-zero process on $\mathcal{V}$.
\begin{condition}
    \label{cond_1:eigenvalue_gap}
    There exists a constant $a_1>1$ such that the eigenvalues of $M(v)$ satisfy
    $\lambda_l-\lambda_{l+1}\ge Cl^{-a_1-1}\text{ for }l\ge1$.
\end{condition}

\begin{condition} \label{cond_2:true_param}
    The true value of $\mbf{\theta}$, denoted by $\mbf{\theta}_0 = [\{\mbf{\gamma}_{jk0}\}_{(j,k) \in \mathcal{D}}, \sigma_0^2 ]$, lies in the interior of a known compact set $\Theta\subset\mathbb{R}^{|\mathcal{D}|d+1}$.
    The true value of $\mathcal{H}$, denoted by $\mathcal{H}_0=\{H_{jk0} \}_{(j,k) \in \mathcal{D}}$, is continuously differentiable with positive derivatives on $[0,\tau]$.
    In addition, for $(j,k)\in\mathcal{D}$, the true value of $\beta_{jk}(v)$, denoted by $\beta_{jk0}(v)=\sum_{l=1}^\infty\beta_{jkl0}\phi_l(v)$,
    satisfies $|\beta_{jkl0}|\le Cl^{-a_2}\mathrm{~for~}l>1$ and some constant $a_2 \ge 1$.
\end{condition}

\begin{condition}
    \label{cond_3:truncation_rate}
    The truncation number $r_n$ satisfies $r_n^{-1} = o(1)$ and $r_n^{2a_1+2a_2+3} = o(n)$.
\end{condition}

\begin{condition}
    \label{cond_4:functional_M1}
    For any $c>0$, $E(e^{c||M||_{L_2}})<\infty$ and $\sup_{v\in\mathcal{V}}E\{|M(v)|^c\}<\infty$, and there exists a $\kappa>0$ such that $\sup_{v_1,v_2\in\mathcal{V}}E[\{|v_1-v_2|^{-\kappa}|M(v_1)-M(v_2)|\}^c]<\infty$.
\end{condition}

\begin{condition}
    \label{cond_5:functional_M2}
    For every integer $k\ge1,\lambda_l^{-k}E\{\int_{\mathcal{V}}M(v)\phi_l(v)dv\}^{2k}$ is bounded uniformly in $l$.
\end{condition}

Condition \ref{cond_1:eigenvalue_gap} is imposed in \cite{hall2007methodology} and controls the decay rate of the eigenvalues $\{\lambda_l\}_{l=1}^{\infty}$.
The first and second parts of Condition \ref{cond_2:true_param} are standard for semiparametric regression models.
The third part assumes that each true functional parameter $\beta_{jk}(v)$ lies in the span of the eigenfunctions $\{\phi_l(v)\}_{l=1}^{\infty}$ and that the associated coefficient sequence has polynomial decay of order $l^{-a_2}$, so that $\sum_{(j,k)\in\mathcal{D}}\sum_{l=1}^{\infty}\beta_{jkl0}^2<\infty$.
Condition \ref{cond_3:truncation_rate} specifies the divergence rate of the truncation number $r_n$.
Conditions \ref{cond_4:functional_M1} and \ref{cond_5:functional_M2} jointly impose strong regularity and moment assumptions on $M(v)$.
Condition \ref{cond_4:functional_M1} requires sub-exponential tails for $||M||_{L_2}$, uniformly bounded finite moments, and H\"older-type control of the increments of $M(\cdot)$ over $\mathcal{V}$.
In particular, the first part implies that $\sum_{l=1}^{\infty}\lambda_l<\infty$.
Condition \ref{cond_5:functional_M2} ensures uniformly bounded higher-order moments for the spectral coefficients of $M$.
Both Conditions \ref{cond_4:functional_M1} and \ref{cond_5:functional_M2} are satisfied when $M(v)$ is a Gaussian process with H\"older-continuous sample paths on a compact domain $\mathcal{V}$.

The second set of conditions is adapted from those used in \citet{gu2024maximum} for semiparametric regression analysis of interval-censored multistate data.

\begin{condition} \label{cond6:X}
    With probability 1, the vector of scalar covariates $\mbf{X}(t)$ is continuously differentiable in $[0, \tau]$. Let $\mbf{\Xi}_{r_n} = (\xi_{1}, \ldots, \xi_{r_n})^{\T}$ be the vector of the first $r_n$ true FPC scores. If there exist a deterministic function $\alpha_1(t)$ and constant vectors $\mbf{\alpha}_2$ and $\mbf{\alpha}_3$ such that $\alpha_1(t)+\mbf{\alpha}_2^{\T}\mbf{X}(t) + \mbf{\alpha}_3^{\T} \mbf{\Xi}_{r_n} = 0$ with probability 1, then $\alpha_1(t)=0$ for $t\in[0,\tau]$, $\mbf{\alpha}_2=\mbf{0}$, and $\mbf{\alpha}_3=\mbf{0}$.
\end{condition}

\begin{condition} \label{cond7:init_state}
    The support of the initial state $S_0$ covers all the nonabsorbing states among $1,\ldots,K$, where an absorbing state is a state that cannot transition to any other state.
\end{condition}

\begin{condition} \label{cond8:exam_time}
    The number of examination times $Q$ is positive with $E(Q)<\infty$. The conditional probability $\mathrm{pr}(\tau_Q=\tau\mid Q,\mbf{X},M)$ is greater than some positive constant $\eta_1$. In addition, the conditional densities of $(\tau_{q-1},\tau_q)$ given $(Q,\mbf{X},M)$, denoted $f_q(t_1,t_2)$ $(q=1,\ldots,Q)$, have continuous second-order partial derivatives with respect to $t_1$ and $t_2$ when $t_2-t_1\ge\eta_2$ for some positive constant $\eta_2$, and are continuously differentiable functionals with respect to $\mbf{X}$ and $M$. Finally, $\operatorname{pr}\{\min_{1\le q\le Q}(\tau_q-\tau_{q-1})\ge\eta_2|Q,\mbf{X},M\}=1$.
\end{condition}

\begin{condition} \label{cond9:identifiability}
    If $(\mbf{\theta}_1,\mathcal{B}_{r_n1},\mathcal{H}_1)$ and $(\mbf{\theta}_2,\mathcal{B}_{r_n2},\mathcal{H}_2)$ satisfy
    $$\int_{b} \mbf{P}(0,t;\mbf{X},\hat{\mbf{\Xi}},b,\mbf{\theta}_1,\mathcal{B}_{r_n1},\mathcal{H}_1) f(b;\sigma_1^2)db
    =\int_{b} \mbf{P}(0,t;\mbf{X},\hat{\mbf{\Xi}},b,\mbf{\theta}_2,\mathcal{B}_{r_n2},\mathcal{H}_2) f(b;\sigma_2^2)db$$
with probability 1 for any $t \in [0, \tau]$, then $\mbf{\theta}_1 = \mbf{\theta}_2$, $\mathcal{B}_{r_n1} = \mathcal{B}_{r_n2}$, and $\mathcal{H}_1(t)=\mathcal{H}_2(t)$ for $t \in [0,\tau]$.
\end{condition}

\begin{condition}\label{cond10:information_invertible}
    Suppose there exist a $K \times K$ matrix-valued function $\mbf{G}(t; b)$ and a scalar constant $c$ such that
\begin{align*}
    \int_{b} \Bigg[ & \int_0^t \mbf{P}(0, s; \mbf{X}, \mbf{\Xi}_{r_n}, b, \mbf{\theta}_0, \mathcal{B}_{r_{n0}}, \mathcal{H}_0) \, \mathrm{d}\mbf{G}(s; b) \, \mbf{P}(s, t; \mbf{X}, \mbf{\Xi}_{r_n}, b, \mbf{\theta}_0, \mathcal{B}_{r_{n0}}, \mathcal{H}_0) \\
    & + \mbf{P}(0, t; \mbf{X}, \mbf{\Xi}_{r_n}, b, \mbf{\theta}_0, \mathcal{B}_{r_{n0}}, \mathcal{H}_0) \frac{c f'(b; \sigma_0^2)}{f(b; \sigma_0^2)} \Bigg] f(b; \sigma_0^2) \, \mathrm{d}b = \mbf{0}
\end{align*}
    with probability 1 for any $t \in [0, \tau]$, where $f'(b; \sigma^2)$ is the derivative of $f(b; \sigma^2)$ with respect to $\sigma^2$, then $\mbf{G}(t; b) = \mbf{0}$ for all $t \in [0, \tau]$ and $c = 0$.
\end{condition}

\subsection{Lemmas}
\label{S1.2:Lemmas}
Recall that the covariance function of $M(v)$ is defined as $\Sigma(v,v^{\prime}) = \text{cov}\{M(v),M(v^{\prime})\}$, and let $\hat{\Sigma}$ denote the corresponding empirical covariance function.
Recall also that $\phi_j$ and $\hat{\phi}_j$ denote the true and estimated $j$-th eigenfunction of this covariance function, respectively.
Define $\hat{\Delta}=(\int_{\mathcal{V}}\int_{\mathcal{V}}|\hat{\Sigma}(v, v^{\prime})-\Sigma(v, v^{\prime})|^{2} dv dv^{\prime} )^{1/2}$ and the minimum consecutive eigenvalue gap up to the $(j+1)$-th eigenvalue as $\delta_j=\min_{1\le k\le j}(\lambda_k-\lambda_{k+1})$, where $k$ and $j$ are positive integers.

\begin{lemma}
    \label{lemma1}
    Assume that with probability 1, $M(v)$ is left-continuous at each $v\in\mathcal{V}$ (or right-continuous at each $v$), and that Conditions 4 and 5 hold. Then, for each integer $k > 0$,
    $$E(\hat{\Delta}^k) < C n^{-k/2}.$$
\end{lemma}

\begin{lemma}
    \label{lemma2}
    Under Conditions 4 and 5, we have:
    $$||\hat{\phi}_j-\phi_j||_{L_2}\le 8^{1/2}\delta_j^{-1} \hat\Delta, \quad\textit{for any positive integer j}.$$
\end{lemma}

\begin{lemma}
    \label{lemma3}
    Let \(\mbf{Z}_1, \dots, \mbf{Z}_n\) be i.i.d. \(r_n \times 1\) random vectors with \(E(\mbf{Z}_i) = \mbf{0}\) and \(E(\mbf{Z}_i \mbf{Z}_i^{\T}) = \mbf{I}_{r_n}\), where \(\mbf{I}_{r_n}\) is the \(r_n \times r_n\) identity matrix. Let \(\tilde{\mbf{Z}}_n = n^{-1/2} \sum_{i=1}^n \mbf{Z}_i\). Assuming \(E\{(\mbf{Z}_i^{\T} \mbf{Z}_i)^2\} = o(n r_n)\), then:
\[
(\tilde{\mbf{Z}}_n^{\T} \tilde{\mbf{Z}}_n - r_n) / (2 r_n)^{1/2} \overset{d}{\to} N(0,1).
\]
\end{lemma}
Lemma~\ref{lemma1} is the same as Lemma 3.3 of \cite{hall2009theory}. Lemma~\ref{lemma2} is the same as Theorem 3 of \cite{hall2006properties}. Lemma~\ref{lemma3} is a special case of Corollary 1 of \cite{peng2018asymptotic}.

\subsection{Proof of Theorem 1}
\label{S1.3:ProofThm1}
We use the following notation throughout: $\mathbb{P}_{n}$ denotes the empirical measure for $n$ independent subjects, $\mathbb{P}$ denotes the true probability measure, and $\mathbb{G}_n = n^{1/2}(\mathbb{P}_n - \mathbb{P})$ is the corresponding empirical process. The notation $a \lesssim b$ means $a \leq Cb$ for some constant $C$.
Recall that the working model parameters consist of the scalar regression parameters $\mbf{\theta} = [\{\mbf{\gamma}_{jk}\}_{(j,k) \in \mathcal{D}}, \sigma^2 ]$, the $r_n$-dimensional FPCA-based imaging coefficients $\mathcal{B}_{r_n} = \{\mbf{B}_{jk}\}_{(j,k) \in \mathcal{D}}$, and the cumulative baseline intensity functions $\mathcal{H}=\{H_{jk} \}_{(j,k) \in \mathcal{D}}$.
To account for the truncation error, we denote by $\mathcal{B}_{+}= \{ \mbf{B}_{jk+}   \}_{ (j,k) \in \mathcal{D} }$ the residual FPCA-based imaging coefficients, where $\mbf{B}_{jk+} = (\beta_{jk,r_n+1 }, \beta_{jk, r_n+2}, \ldots)^{\T}$ is the transition-specific, infinite-dimensional vector of the residual imaging coefficients beyond the truncation level $r_n$.
Accordingly, let $\mbf{\Xi} = (\xi_{1}, \ldots, \xi_{r_n}, \ldots)^{\T}$ be the infinite-dimensional vector of the true FPC scores. We use $\hat{\mbf{\Xi}}$ to denote the estimated $r_n$-truncated FPC scores obtained from the FPCA step.
The likelihood function for a single subject is then defined as: $L(\mbf{\theta}, \mathcal{B}_{r_n}, \mathcal{B}_{+},\mathcal{H}; \mbf{\Xi})= \int_{b}\prod_{q=1}^{Q}\mbf{P}(\tau_{q-1},\tau_{q};\mbf{X},\mbf{\Xi},b)^{(S_{q-1},S_q)}f(b;\sigma^2)db$ with the corresponding log-likelihood denoted by $\ell(\mbf{\theta}, \mathcal{B}_{r_n}, \mathcal{B}_{+},\mathcal{H}; \mbf{\Xi}) = \log L(\mbf{\theta}, \mathcal{B}_{r_n}, \mathcal{B}_{+},\mathcal{H}; \mbf{\Xi})$.

To handle the estimation error and FPCA approximation error, we distinguish among the following three sets of parameters:
\begin{enumerate}
    \item The proposed estimators $(\hat{\mbf{\theta}}, \hat{\mathcal{B}}_{r_n} ,\hat{\mathcal{H}})$ that maximize the empirical log pseudo likelihood under the working model with truncated, estimated FPC scores:
        $$(\hat{\mbf{\theta}}, \hat{\mathcal{B}}_{r_n} ,\hat{\mathcal{H}}) = \arg \max \mathbb{P}_n \ell(\mbf{\theta}, \mathcal{B}_{r_n}, \mbf{0}, \mathcal{H}; \hat{\mbf{\Xi}})$$
    \item The pseudo-true parameters $({\mbf{\theta}}^*, {\mathcal{B}^*_{r_n}} ,{\mathcal{H}}^* )$ that maximize the expected log pseudo likelihood under the working model with truncated, estimated FPC scores:
        $$({\mbf{\theta}}^*, {\mathcal{B}^*_{r_n}} ,{\mathcal{H}}^* ) = \arg \max \mathbb{P} \ell(\mbf{\theta}, \mathcal{B}_{r_n}, \mbf{0}, \mathcal{H}; \hat{\mbf{\Xi}}),$$
        where $\hat{\mbf{\Xi}}$ is treated as fixed covariates when taking the expectation.
    \item The true parameters $({\mbf{\theta}}_0, {\mathcal{B}_{r_n0}} ,{\mathcal{H}}_0 )$ that maximize the expected log-likelihood under the true underlying model:
        $$({\mbf{\theta}}_0, {\mathcal{B}_{r_n0}} ,{\mathcal{H}}_0 ) =  \arg \max \mathbb{P} \ell(\mbf{\theta}, \mathcal{B}_{r_n}, \mathcal{B}_{+0}, \mathcal{H}; {\mbf{\Xi}} ),$$
        where $\mathcal{B}_{+0}$ denotes the true value of $\mathcal{B}_{+}$.
\end{enumerate}

To establish Theorem 1, we proceed in three main steps. First, we show that the proposed estimators $(\hat{\mbf{\theta}}, \hat{\mathcal{B}}_{r_n} ,\hat{\mathcal{H}})$ are consistent for the pseudo-true parameters $ ( \mbf{\theta}^*, \mbf{\mathcal{B}}^*_{r_n} , \mathcal{H}^*)$. Second, we quantify the discrepancy between $({\mbf{\theta}}^*, {\mathcal{B}^*_{r_n}} ,{\mathcal{H}}^* )$ and the true parameters $({\mbf{\theta}}_0, {\mathcal{B}_{r_n0}} ,{\mathcal{H}}_0 )$. Finally, we extend to derive the convergence rate for the functional coefficient estimators $\hat{\beta}_{jk}(v)$.

{\it Step 1.} We prove that the $(\hat{\mbf{\theta}}, \hat{\mathcal{B}}_{r_n} ,\hat{\mathcal{H}})$ converges to $({\mbf{\theta}}^*, {\mathcal{B}^*_{r_n}} ,{\mathcal{H}}^* ) $ in probability.
Since our Conditions \ref{cond_2:true_param} and \ref{cond6:X}--\ref{cond10:information_invertible} satisfy all the conditions required in \cite{gu2024maximum}, we can directly invoke Lemma 2 of \cite{gu2024maximum} to obtain the following convergence rate:
\[
\|\hat{\mbf{\theta}} - \mbf{\theta}^*\| + \| \hat{\mathcal{B}}_{r_n} - \mathcal{B}^{*}_{r_n}  \| + \|\hat{\mathcal{H}} - \mathcal{H}^*\|_{L_2} = O_p(n^{-1/3}).
\]

{\it Step 2.} We characterize the approximation error between the pseudo-true parameters $( \mbf{\theta}^*, \mbf{\mathcal{B}}^*_{r_n} , \mathcal{H}^*)$ and the true parameters $({\mbf{\theta}}_0, {\mathcal{B}_{r_n0}} ,{\mathcal{H}}_0 )$. Specifically, for any fixed $\epsilon \in (0, a_2/2)$, define $\alpha_n=O_p(r_n^{a_1/2+3/4+\epsilon}n^{-1/4}+r_n^{-a_2/2+\epsilon})$. We aim to show that
\[
\|{\mbf{\theta}}^* - \mbf{\theta}_0\| +  \| {\mathcal{B}}^{*}_{r_n} - \mathcal{B}_{r_n0}  \| + \|\mathcal{H}^* - \mathcal{H}_0\|_{L_2}  = O_p(\alpha_n).
\]
To establish this convergence rate, it suffices to show that, with probability 1,
$$
\sup_{({\mbf{\theta}}, {\mathcal{B}_{r_n}} ,{\mathcal{H}} ) \in \partial N_{\alpha_n}(\mbf{\theta}_0, {\mathcal{B}_{r_n0}} ,{\mathcal{H}}_0  )} \mathbb{P}  \ell(\mbf{\theta}, \mathcal{B}_{r_n}, 0,\mathcal{H}; \hat{\mbf{\Xi}} ) < \mathbb{P} \ell(\mbf{\theta}_0, \mathcal{B}_{r_n0}, \mbf{0}, \mathcal{H}_0; \hat{\mbf{\Xi}} )
$$
where $\partial N_{\alpha_n}(\mbf{\theta}_0, {\mathcal{B}_{r_n0}} ,{\mathcal{H}}_0  )$ denotes the boundary of a neighborhood of radius \(\alpha_n\) around the true parameters $(\mbf{\theta}_0, {\mathcal{B}_{r_n0}} ,{\mathcal{H}}_0  )$.
By telescope method, we can rewrite the difference:
    \begin{align}
        D&= \mathbb{P} \ell(\mbf{\theta}, \mathcal{B}_{r_n}, \mbf{0},\mathcal{H}; \hat{\mbf{\Xi}} ) - \mathbb{P} \ell(\mbf{\theta}_0, \mathcal{B}_{r_n0}, \mbf{0}, \mathcal{H}_0; \hat{\mbf{\Xi}}) \label{D_total} \\
        &= \mathbb{P} \ell(\mbf{\theta}, \mathcal{B}_{r_n}, \mbf{0},\mathcal{H}; \hat{\mbf{\Xi}} ) - \mathbb{P} \ell(\mbf{\theta}, \mathcal{B}_{r_n}, \mathcal{B}_{+0}, \mathcal{H}; \mbf{\Xi} ) \label{D1} \\
        &+ \mathbb{P} \ell(\mbf{\theta}, \mathcal{B}_{r_n}, \mathcal{B}_{+0},\mathcal{H}; \mbf{\Xi} ) - \mathbb{P} \ell(\mbf{\theta}_0, \mathcal{B}_{r_n0}, \mathcal{B}_{+0},\mathcal{H}_0; \mbf{\Xi}) \label{D2} \\
        &+ \mathbb{P} \ell(\mbf{\theta}_0, \mathcal{B}_{r_n0}, \mathcal{B}_{+0},\mathcal{H}_0; \mbf{\Xi} ) - \mathbb{P} \ell(\mbf{\theta}_0, \mathcal{B}_{r_n0}, \mbf{0}, \mathcal{H}_0; \hat{\mbf{\Xi}} )  \label{D3}
    \end{align}
Let $\mbf{v} = (\mbf{\theta}-\mbf{\theta}_{0}, \mathcal{B}_{r_n} - \mathcal{B}_{r_n0} , \mathcal{H}-\mathcal{H}_0)$ denote the perturbation direction, and let $\|\mbf{v}\|^2 = \|\mbf{\theta}- \mbf{\theta}_0\|^2 + \| \mathcal{B}_{r_n} - \mathcal{B}_{r_n0} \|^2 + \|\mathcal{H}-\mathcal{H}_0\|_{L_2}^2$. For $({\mbf{\theta}}, {\mathcal{B}_{r_n}} ,{\mathcal{H}} ) \in \partial N_{\alpha_n}(\mbf{\theta}_0, {\mathcal{B}_{r_n0}} ,{\mathcal{H}}_0)$, we have $\|\mbf{v}\| = \alpha_n$. We apply Taylor's expansion for \eqref{D2} around the true parameter \((\mbf{\theta}_{0}, \mathcal{B}_{r_n0}, \mathcal{H}_0)\). Since $\mathbb{P} \dot{\ell}(\mbf{\theta}_0, \mathcal{B}_{r_n0}, \mathcal{B}_{+0},\mathcal{H}_0; \mbf{\Xi} )[\mbf{v}] = 0$ by the definition of the maximizer, the first-order term vanishes.
The second-order term gives:
\begin{align*}
        \eqref{D2} &=  \frac{1}{2} \mathbb{P} \ddot{\ell}( \mbf{\theta}_0, \mathcal{B}_{r_n0}, \mathcal{B}_{+0},\mathcal{H}_0; \mbf{\Xi} ) [\mbf{v}, \mbf{v}] + o(\|\mbf{v}\|^2)\\
        &=-  \frac{1}{2}\mathbb{P}\dot{\ell}^{2}(\mbf{\theta}_0, \mathcal{B}_{r_n0}, \mathcal{B}_{+0},\mathcal{H}_0; \mbf{\Xi}) [\mbf{v}] + o(\|\mbf{v}\|^2)\\
        & \lesssim - \|\mbf{v}\|^2 =-\alpha_n^2
\end{align*}
where the last inequality follows from similar arguments to the proof of Lemma 3 of \citet{gu2025semiparametric}.

Next, we consider \eqref{D1} and \eqref{D3} together:
    $$
    \left| \eqref{D1} + \eqref{D3} \right|
    \leq 2 \sup_{({\mbf{\theta}}, {\mathcal{B}_{r_n}} ,{\mathcal{H}} ) \in \partial N_{\alpha_n}(\mbf{\theta}_0, {\mathcal{B}_{r_n0}} ,{\mathcal{H}}_0  )} \left| \mathbb{P} \ell(\mbf{\theta}, \mathcal{B}_{r_n}, \mbf{0} ,\mathcal{H}; \hat{\mbf{\Xi}} )  - \mathbb{P} \ell(\mbf{\theta}, \mathcal{B}_{r_n}, \mathcal{B}_{+0}, \mathcal{H}; \mbf{\Xi})   \right|
    $$
By the mean value theorem, we have:
\begin{align*}
    & \ell(\mbf{\theta}, \mathcal{B}_{r_n}, \mbf{0},\mathcal{H}; \hat{\mbf{\Xi}} ) - \ell(\mbf{\theta}, \mathcal{B}_{r_n}, \mathcal{B}_{+0}, \mathcal{H}; \mbf{\Xi} ) \\
    &= \frac{1}{L(\mbf{\theta}, \mathcal{B}_{r_n}, \bar{\mathcal{B}}_{+},\mathcal{H}; \bar{\mbf{\Xi}})} \int_{b} \sum_{q=1}^{Q} \Bigg\{ \prod_{q' \neq q} \mbf{P}(\tau_{q'-1}, \tau_{q'}; \mbf{X}, \bar{\mbf{\Xi}}, b, \bar{\mathcal{B}}_{+})^{(S_{q'-1}, S_{q'})} \Bigg\} \\
    &\quad \times \Big\{ \mbf{P}(\tau_{q-1}, \tau_{q}; \mbf{X}, \hat{\mbf{\Xi}}, b, \mbf{0}) - \mbf{P}(\tau_{q-1}, \tau_{q}; \mbf{X}, \mbf{\Xi}, b, \mathcal{B}_{+0}) \Big\}^{(S_{q-1}, S_q)} f(b; \sigma^2) db,
\end{align*}
where the intermediate values $(\bar{\mbf{\Xi}}, \bar{\mathcal{B}}_+)$ lie between $(\hat{\mbf{\Xi}}, \mbf{0})$ and $(\mbf{\Xi}, \mathcal{B}_{+0})$.
The difference in the transition probability matrices can be expressed using Duhamel's equation in Theorem II.6.2 of \cite{andersen1993statistical}:
\begin{align*}
    & \mbf{P}(\tau_{q-1}, \tau_{q}; \mbf{X}, \hat{\mbf{\Xi}}, b, \mbf{0}) - \mbf{P}(\tau_{q-1}, \tau_{q}; \mbf{X}, \mbf{\Xi}, b, \mathcal{B}_{+0}) \\
    &= \int_{\tau_{q-1}}^{\tau_{q}} \mbf{P}(\tau_{q-1}, t; \mbf{X}, \hat{\mbf{\Xi}}, \mbf{0}, b, \mbf{0})
    \times \Big\{ d\mbf{A}(t; \mbf{X}, \hat{\mbf{\Xi}}, b, \mbf{0}) - d\mbf{A}(t; \mbf{X}, \mbf{\Xi}, b, \mathcal{B}_{+0}) \Big\}
    \times \mbf{P}(t, \tau_{q}; \mbf{X}, \mbf{\Xi}, b, \mathcal{B}_{+0}),
\end{align*}
where $\mbf{A}(t; \cdot)$ is the cumulative transition intensity matrix for a subject evaluated at time $t$.

To streamline the expression, we define
    \begin{gather*}
    z_{jk}(t) = \mbf{\gamma}_{jk}^{\T} \mbf{X}(t) + \sum_{l=1}^{r_n} \beta_{jkl} \xi_l + \sum_{l=r_n+1}^{\infty} \beta_{jkl0} \xi_l   +b, \\
    \epsilon_{jk} = \sum_{l=1}^{r_n} \beta_{jkl}(\hat \xi_l - \xi_l) - \sum_{l=r_n+1}^{\infty} \beta_{jkl0} \xi_l,
    \end{gather*}
where $z_{jk}(t)$ represents the true and full predictors, and $\epsilon_{jk}$ represents the approximation error induced by truncated FPCA.
The difference in the transition intensity functions is given by
\begin{align*}
    & d A_{jk}(t; \mbf{X}, \hat{\mbf{\Xi}}, b, \mbf{0}) - d A_{jk}(t; \mbf{X}, \mbf{\Xi}, b, \mathcal{B}_{+0}) \\
    &= \Bigg[ \exp\Bigg\{ \mbf{\gamma}_{jk}^{\T} \mbf{X}(t) + \sum_{l=1}^{r_n} \beta_{jkl}\hat{\xi}_{l} + b \Bigg\} - \exp\Bigg\{ \mbf{\gamma}_{jk}^{\T} \mbf{X}(t) + \sum_{l=1}^{r_n} \beta_{jkl}\xi_{l} + \sum_{l=r_n+1}^{\infty} \beta_{jkl0}\xi_{l} + b \Bigg\} \Bigg] d H_{jk}(t) \\
    &= \Big[ \exp\big\{ z_{jk}(t) + \epsilon_{jk} \big\} - \exp\big\{ z_{jk}(t) \big\} \Big] d H_{jk}(t) \\
    &= \exp\big\{ z_{jk}(t) + \epsilon_{jk}^* \big\} \epsilon_{jk} d H_{jk}(t),
\end{align*}
where $\epsilon_{jk}^*$ lies between $0$ and $\epsilon_{jk}$.
Substituting this difference into the absolute value of the transition probability difference, we obtain:
\begin{align*}
    &\left| \mbf{P}(\tau_{q-1}, \tau_q; \mbf{X}, \hat{\mbf{\Xi}}, b, \mbf{0})^{(S_{q-1}, S_q)} - \mbf{P}(\tau_{q-1}, \tau_q; \mbf{X}, \mbf{\Xi}, b, \mathcal{B}_{+0})^{(S_{q-1}, S_q)} \right| \\
    &= \Bigg| \int_{\tau_{q-1}}^{\tau_q} \Big\{ \mbf{P}(\tau_{q-1}, t; \mbf{X}, \hat{\mbf{\Xi}}, b, \mbf{0}) \big[ d\mbf{A}(t; \mbf{X}, \hat{\mbf{\Xi}}, b, \mbf{0}) \\
    &\quad - d\mbf{A}(t; \mbf{X}, \mbf{\Xi}, b, \mathcal{B}_{+0}) \big] \mbf{P}(t, \tau_q; \mbf{X}, \mbf{\Xi}, b, \mathcal{B}_{+0}) \Big\}^{(S_{q-1}, S_q)} \Bigg| \\
    &= \Bigg| \int_{\tau_{q-1}}^{\tau_q} \sum_{(j,k) \in \mathcal{D}} \mbf{P}(\tau_{q-1}, t; \mbf{X}, \hat{\mbf{\Xi}}, b, \mbf{0})^{(S_{q-1}, j)} \Big[ \exp\big\{ z_{jk}(t) + \epsilon_{jk}^* \big\} \epsilon_{jk} \Big] dH_{jk}(t) \\
    &\quad \times \Big\{ \mbf{P}(t, \tau_q; \mbf{X}, \mbf{\Xi}, b, \mathcal{B}_{+0})^{(k, S_q)} - \mbf{P}(t, \tau_q; \mbf{X}, \mbf{\Xi}, b, \mathcal{B}_{+0})^{(j, S_q)} \Big\} \Bigg| \\
    &\le \int_{\tau_{q-1}}^{\tau_q} \sum_{(j,k) \in \mathcal{D}} \exp\big\{ z_{jk}(t) + \epsilon_{jk}^* \big\} |\epsilon_{jk}| dH_{jk}(t),
\end{align*}
where the second equality follows from expanding the matrix product and using the fact that the row sums of the transition intensity matrices are zero, and the last inequality follows from the triangle inequality along with the fact that transition probabilities are bounded between $0$ and $1$ (i.e., $| \mbf{P}^{(k, S_q)} - \mbf{P}^{(j, S_q)} | \le 1$ and $\mbf{P}^{(S_{q-1}, j)} \le 1$).

Then, by applying the Cauchy-Schwarz inequality, we have:
\begin{align*}
    &\left| \mathbb{P} \Big[ \ell(\mbf{\theta}, \mathcal{B}_{r_n}, \mbf{0}, \mathcal{H}; \hat{\mbf{\Xi}}) - \ell(\mbf{\theta}, \mathcal{B}_{r_n}, \mathcal{B}_{+0}, \mathcal{H}; \mbf{\Xi}) \Big] \right| \\
    &\leq \mathbb{P} \left| \ell(\mbf{\theta}, \mathcal{B}_{r_n}, \mbf{0}, \mathcal{H}; \hat{\mbf{\Xi}}) - \ell(\mbf{\theta}, \mathcal{B}_{r_n}, \mathcal{B}_{+0}, \mathcal{H}; \mbf{\Xi}) \right| \\
    &\leq \mathbb{P} \Bigg[ \frac{1}{L(\mbf{\theta}, \mathcal{B}_{r_n}, \bar{\mathcal{B}}_{+}, \mathcal{H}; \bar{\mbf{\Xi}} )} \int_{b} \sum_{q=1}^{Q} \Bigg\{ \prod_{q' \neq q} \mbf{P}(\tau_{q'-1}, \tau_{q'}; \mbf{X}, \bar{\mbf{\Xi}}, b, \bar{\mathcal{B}}_{+})^{(S_{q'-1}, S_{q'})} \Bigg\} \\
    &\quad \times \left| \mbf{P}(\tau_{q-1}, \tau_{q}; \mbf{X}, \hat{\mbf{\Xi}}, b, \mbf{0})^{(S_{q-1}, S_{q})} - \mbf{P}(\tau_{q-1}, \tau_{q}; \mbf{X}, \mbf{\Xi}, b, \mathcal{B}_{+0})^{(S_{q-1}, S_{q})} \right| f(b;\sigma^2) db \Bigg] \\
    &\leq \mathbb{P} \Bigg[ \frac{1}{L(\mbf{\theta}, \mathcal{B}_{r_n}, \bar{\mathcal{B}}_{+}, \mathcal{H}; \bar{\mbf{\Xi}})} \int_{b} \int_{0}^{\tau} \sum_{(j,k) \in \mathcal{D}} \exp\big\{ z_{jk}(t) + \epsilon_{jk}^* \big\} |\epsilon_{jk}| dH_{jk}(t) f(b;\sigma^2) db \Bigg] \\
    &\leq \mathbb{P} \Bigg[ \frac{1}{L(\mbf{\theta}, \mathcal{B}_{r_n}, \bar{\mathcal{B}}_{+}, \mathcal{H}; \bar{\mbf{\Xi}})} \int_{b} \int_{0}^{\tau} \sum_{(j,k) \in \mathcal{D}} \exp\big\{ z_{jk}(t) + \epsilon_{jk}^* \big\} \\
    &\quad \times \Bigg\{ \left| \sum_{l=1}^{r_n} \beta_{jkl} (\hat{\xi}_{l} - \xi_{l}) \right| + \left| \sum_{l=r_n+1}^{\infty} \beta_{jkl0} \xi_{l} \right| \Bigg\} dH_{jk}(t) f(b;\sigma^2) db \Bigg],
\end{align*}
where the third inequality leverages the fact that the transition probabilities are bounded by 1 (i.e., $\prod_{q' \neq q} \mbf{P}^{(S_{q'-1}, S_{q'})} \le 1$).

By Conditions \ref{cond_2:true_param} and \ref{cond6:X}, we have $|\mbf{\gamma}_{jk}^{\T} \mbf{X}(t)| \le \|\mbf{\gamma}_{jk}\| \|\mbf{X}(t)\| \le C$ for every $(j,k) \in \mathcal{D}$.
For the functional components, since $M(v)$ is a mean-zero process, we can obtain
\(
    \|M\|_{L_2}^2 = \int_{\mathcal{V}} \left\{ \sum_{l=1}^{\infty} \xi_{l}\phi_l(v) \right\}^2 dv = \sum_{l=1}^{\infty}\left\{ \xi_{l}^2 \int_{\mathcal{V}} \phi_l^2(v)dv \right\} = \sum_{l=1}^{\infty} \xi_{l}^2.
\)
Similarly, for the estimated scores $\hat{\xi}_l = \int_{\mathcal{V}} M(v)\hat{\phi}_l(v)dv$, since the estimated eigenfunctions $\{\hat{\phi}_l\}_{l=1}^{r_n}$ also form an orthonormal system, the non-negativity of the squared $L_2$-norm of the residual implies
$
    0 \le \int_{\mathcal{V}} \{ M(v) - \sum_{l=1}^{r_n} \hat{\xi}_l \hat{\phi}_l(v) \}^2 dv = \|M\|_{L_2}^2 - 2\sum_{l=1}^{r_n} \hat{\xi}_l \int_{\mathcal{V}} M(v)\hat{\phi}_l(v)dv + \sum_{l=1}^{r_n} \hat{\xi}_l^2 = \|M\|_{L_2}^2 - \sum_{l=1}^{r_n} \hat{\xi}_l^2,
$
which explicitly ensures $\sum_{l=1}^{r_n} \hat{\xi}_{l}^2 \le \|M\|_{L_2}^2$.
Applying the Cauchy-Schwarz inequality, and bounding the estimated coefficients within the $\alpha_n$-neighborhood by $(\sum_{l=1}^{r_n} \beta_{jkl}^2)^{1/2} \le \alpha_n + \|\beta_{jk0}\|_{L_2}$, for every $(j,k) \in \mathcal{D}$, we obtain:
\begin{align*}
    \left| \sum_{l=1}^{r_n} \beta_{jkl} \xi_{l} + \sum_{l=r_n+1}^{\infty} \beta_{jkl0} \xi_{l} \right| 
    &\le \left( \sum_{l=1}^{r_n} \beta_{jkl}^2 \right)^{1/2} \left( \sum_{l=1}^{r_n} \xi_{l}^2 \right)^{1/2} + \left( \sum_{l=r_n+1}^{\infty} \beta_{jkl0}^2 \right)^{1/2} \left( \sum_{l=r_n+1}^{\infty} \xi_{l}^2 \right)^{1/2} \\
    &\le \big( \alpha_n + \|\beta_{jk0}\|_{L_2} \big) \|M\|_{L_2} + \|\beta_{jk0}\|_{L_2} \|M\|_{L_2} \le C \|M\|_{L_2}, \\
    \left| \sum_{l=1}^{r_n} \beta_{jkl} \hat{\xi}_{l} \right| 
    &\le \left( \sum_{l=1}^{r_n} \beta_{jkl}^2 \right)^{1/2} \left( \sum_{l=1}^{r_n} \hat{\xi}_{l}^2 \right)^{1/2} \le \big( \alpha_n + \|\beta_{jk0}\|_{L_2} \big) \|M\|_{L_2} \le C \|M\|_{L_2},
\end{align*}
where we use the fact that the true coefficients satisfy $\sum_{l=1}^{\infty} \beta_{jkl0}^2 < \infty$ and that $\alpha_n = o(1)$ under Condition~\ref{cond_3:truncation_rate}.
Since $\epsilon_{jk}^*$ lies between $0$ and $\epsilon_{jk}$, both $z_{jk}(t)$ and $z_{jk}(t) + \epsilon_{jk}$ are bounded by $C + C\|M\|_{L_2} + |b|$, which naturally implies $z_{jk}(t) + \epsilon_{jk}^* \le C + C\|M\|_{L_2} + |b|$. Given that $|\mathcal{D}|$ is finite and $\{H_{jk}\}_{(j,k)\in\mathcal{D}}$ are continuously differentiable on $[0,\tau]$, the expected integral is bounded by:
\begin{align*}
    &\mathbb{P} \Bigg[ \int_{0}^{\tau} \sum_{(j,k) \in \mathcal{D}} \exp\big\{ z_{jk}(t) + \epsilon_{jk}^* \big\} dH_{jk}(t) \Bigg] 
     = \mathbb{P} \Bigg[ \sum_{(j,k) \in \mathcal{D}} \int_{0}^{\tau} \exp\big\{ z_{jk}(t) + \epsilon_{jk}^* \big\} dH_{jk}(t) \Bigg] \\
    &\le \mathbb{P} \Bigg[ \exp\big( C + C\|M\|_{L_2} + |b| \big) \sum_{(j,k) \in \mathcal{D}} \int_{0}^{\tau} dH_{jk}(t) \Bigg]  \lesssim \mathbb{P} \Big[ \exp\big( C\|M\|_{L_2} \big) \Big] \times \mathbb{P} \Big[ \exp(|b|) \Big] = O(1),
\end{align*}
where the final equality follows from the sub-exponential tail of $\|M\|_{L_2}$ in Condition \ref{cond_4:functional_M1} (i.e., $E(e^{c\|M\|_{L_2}}) < \infty$) and the bounded exponential moments of the random effect $b \sim N(0,\sigma^2)$.


By the properties of FPCA, the FPC scores $\xi_{l}$ have mean zero and are mutually uncorrelated with $\mathrm{var}(\xi_{l}) = \lambda_l$. Therefore, we can obtain 
$
E(\sum_{l=r_n+1}^{\infty} \beta_{jkl0} \xi_l)^2=\mathrm{var}(\sum_{l=r_n+1}^{\infty} \beta_{jkl0} \xi_l) = \sum_{l=r_n+1}^{\infty} \beta_{jkl0}^2 \lambda_l 
$.
By Condition \ref{cond_2:true_param}, the true functional coefficients satisfy $|\beta_{jkl0}| \le C l^{-a_2}$. For the tail indices $l \ge r_n+1$, we can uniformly bound these coefficients by their maximum $\beta_{jkl0}^2 \le C^2 (r_n+1)^{-2a_2} \lesssim r_n^{-2a_2}$, yielding $    \sum_{l=r_n+1}^{\infty} \beta_{jkl0}^2 \lambda_l 
    \lesssim r_n^{-2a_2} \sum_{l=r_n+1}^{\infty} \lambda_l$.
Furthermore, Condition \ref{cond_4:functional_M1} guarantees that  $\sum_{l=1}^{\infty} \lambda_l < \infty$ implying $\sum_{l=r_n+1}^{\infty} \lambda_l = o(1)$. Substituting this convergence rate back into the inequality yields the final bound:
\begin{align*}
    E\Bigg( \sum_{l=r_n+1}^{\infty} \beta_{jkl0} \xi_{l} \Bigg)^2 
    &\lesssim r_n^{-2a_2} \times o(1) = o(r_n^{-2a_2}).
\end{align*}

Prior to deriving the following bound, note that Condition \ref{cond_1:eigenvalue_gap} implies $\delta_l = \min_{1 \le k \le l}(\lambda_k - \lambda_{k+1}) \gtrsim l^{-a_1-1}$. Bounding the finite sum by an integral directly yields $\sum_{l=1}^{r_n} \delta_l^{-2} \lesssim \sum_{l=1}^{r_n} l^{2a_1+2} \lesssim \int_{1}^{r_n+1} x^{2a_1+2} dx = O(r_n^{2a_1+3})$.
To bound the expected squared error of the functional component, we apply the Cauchy-Schwarz inequality sequentially:
\begin{align*} 
    E\Bigg\{ \sum_{l=1}^{r_n} \beta_{jkl}(\hat{\xi}_l - \xi_{l}) \Bigg\}^2 
    &\overset{\text{(i)}}{\le} E\Bigg\{ \Bigg( \sum_{l=1}^{r_n} \beta_{jkl}^2 \Bigg) \sum_{l=1}^{r_n} (\hat{\xi}_l - \xi_{l})^2 \Bigg\} \\ 
    &\overset{\text{(ii)}}{\le} C \cdot E\left[ \sum_{l=1}^{r_n} \Bigg( \int_{\mathcal{V}} M(v)\big\{ \hat{\phi}_l(v) - \phi_l(v) \big\} dv \Bigg)^2 \right] \\ 
    &\overset{\text{(iii)}}{\le} C \cdot E\big( \|M\|_{L_2}^2 \hat{\Delta}^2 \big) \sum_{l=1}^{r_n} 8 \delta_l^{-2} \\ 
    &\overset{\text{(iv)}}{\lesssim} r_n^{2a_1+3} \sqrt{ E\big( \|M\|_{L_2}^4 \big) } \sqrt{ E\big( \hat{\Delta}^4 \big) } \\ 
    &\overset{\text{(v)}}{=} O(r_n^{2a_1+3} n^{-1}), 
\end{align*}
where (i) uses the Cauchy-Schwarz inequality for finite sums; (ii) applies the deterministic bound $\sum_{l=1}^{r_n} \beta_{jkl}^2 \le C$ within the $\alpha_n$-neighborhood and the definition of FPC scores; (iii) follows from the Cauchy-Schwarz inequality for integrals and Lemma \ref{lemma2} ($\|\hat{\phi}_l - \phi_l\|_{L_2}^2 \le 8\delta_l^{-2}\hat{\Delta}^2$); (iv) decouples the expectation via Cauchy-Schwarz inequality and applies $\sum_{l=1}^{r_n} \delta_l^{-2} = O(r_n^{2a_1+3})$; and (v) holds because $E(\|M\|_{L_2}^4) < \infty$ by Condition 4, and $E(\hat{\Delta}^4) = O(n^{-2})$ by Lemma \ref{lemma1}.

Now, we can bound the expected log-likelihood difference as follows:
\begin{align*}
    &\left| \mathbb{P} \Big[ \ell(\mbf{\theta}, \mathcal{B}_{r_n}, \mbf{0}, \mathcal{H}; \hat{\mbf{\Xi}}) - \ell(\mbf{\theta}, \mathcal{B}_{r_n}, \mathcal{B}_{+0}, \mathcal{H}; \mbf{\Xi}) \Big] \right| \\
    &\le \mathbb{P} \Bigg[ \frac{1}{L(\mbf{\theta}, \mathcal{B}_{r_n}, \bar{\mathcal{B}}_{+}, \mathcal{H}; \bar{\mbf{\Xi}})} \int_{b} \int_{0}^{\tau} \sum_{(j,k) \in \mathcal{D}} \exp\big\{ z_{jk}(t) + \epsilon_{jk}^* \big\} \\
    &\quad \times \Bigg\{ \left| \sum_{l=1}^{r_n} \beta_{jkl} (\hat{\xi}_{l} - \xi_{l}) \right| + \left| \sum_{l=r_n+1}^{\infty} \beta_{jkl0} \xi_{l} \right| \Bigg\} dH_{jk}(t) f(b;\sigma^2) db \Bigg] \\
    &\lesssim \mathbb{P} \Bigg[ \int_{b} \exp\big(C + C|b| + C\|M\|_{L_2}\big) \Bigg\{ \left| \sum_{l=1}^{r_n} \beta_{jkl} (\hat{\xi}_{l} - \xi_{l}) \right| + \left| \sum_{l=r_n+1}^{\infty} \beta_{jkl0} \xi_{l} \right| \Bigg\} f(b;\sigma^2) db \Bigg] \\
    &\le \sqrt{ E\Big[ \exp\big(2C + 2C|b| + 2C\|M\|_{L_2}\big) \Big] } \times \sqrt{ E\Bigg( \left| \sum_{l=1}^{r_n} \beta_{jkl}(\hat{\xi}_l - \xi_{l}) \right| + \left| \sum_{l=r_n+1}^{\infty} \beta_{jkl0} \xi_{l} \right| \Bigg)^2 } \\
    &\lesssim \sqrt{ 2E\Bigg\{ \Bigg( \sum_{l=1}^{r_n} \beta_{jkl}(\hat{\xi}_l - \xi_{l}) \Bigg)^2 + \Bigg( \sum_{l=r_n+1}^{\infty} \beta_{jkl0} \xi_{l} \Bigg)^2 \Bigg\} } \\
    &= O\big(r_n^{a_1+3/2}n^{-1/2} + r_n^{-a_2}\big),
\end{align*}
where the second inequality holds because the intermediate values $(\mbf{\theta}, \mathcal{B}_{r_n}, \bar{\mathcal{B}}_{+}, \mathcal{H}; \bar{\mbf{\Xi}})$ lie within the shrinking $\alpha_n$-neighborhood of the true parameters, ensuring the likelihood is bounded away from $0$; the third inequality applies the Cauchy-Schwarz inequality for expectations, and the final inequality substitutes the previously derived bounds $O(r_n^{2a_1+3}n^{-1})$ and $O(r_n^{-2a_2})$ into the expected squared terms. 

By integrating the aforementioned results, the difference $D$ in \eqref{D_total} satisfies the following inequality on the boundary of the $\alpha_n$-neighborhood:
\begin{equation*}
    D \le -C\alpha_n^2 + O\big(r_n^{a_1+3/2}n^{-1/2} + r_n^{-a_2}\big).
\end{equation*}
The right-hand side is strictly negative with probability one. Thus, we conclude that $({\mbf{\theta}}^*, {\mathcal{B}^*_{r_n}} ,{\mathcal{H}}^* )$ lies strictly within the interior of the $\alpha_n$-neighborhood of $({\mbf{\theta}}_0, {\mathcal{B}_{r_n0}} ,{\mathcal{H}}_0 )$, i.e.,
\begin{equation*}
    \|\mbf{\theta}_{r_n}^* - \mbf{\theta}_{0}\| + \| \mathcal{B}_{r_n}^* - \mathcal{B}_{r_n0} \|+ \|\mathcal{H}^* - \mathcal{H}_0\|_{L_2} = O_p(\alpha_n).
\end{equation*}
Finally, we apply the triangle inequality combining the estimation and truncation errors to yield the convergence rate for our proposed estimators.
\begin{align*}
    &\|\hat{\mbf{\theta}} - \mbf{\theta}_{0}\| + \| \hat{\mathcal{B}}_{r_n} - \mathcal{B}_{r_n0} \| + \|\hat{\mathcal{H}} - \mathcal{H}_0\|_{L_2} 
    \\
    & \leq  \big(\|\hat{\mbf{\theta}} - \mbf{\theta}^*\| + \| \hat{\mathcal{B}}_{r_n} - \mathcal{B}^{*}_{r_n}  \| + \|\hat{\mathcal{H}} - \mathcal{H}^*\|_{L_2} \big) + \big( \|\mbf{\theta}_{r_n}^* - \mbf{\theta}_{0}\| + \| \mathcal{B}_{r_n}^* - \mathcal{B}_{r_n0} \|+ \|\mathcal{H}^* - \mathcal{H}_0\|_{L_2} \big)
    \\
    &= O_p(n^{-1/3}) + O_p(\alpha_n)= O_p(\alpha_n) = o_p(1).
\end{align*}

{\it Step 3.}
We establish the convergence rate for $\hat{\mbf{\beta}}_{jk}(v)$:
$$\begin{aligned} \|\hat{\beta}_{jk}(v) - \beta_{jk0}(v)\|_{L_2} & =\| \sum_{l=1}^{r_n}\hat\beta_{jkl}\hat\phi_l(v) - \sum_{l=1}^{\infty} \beta_{jkl0}\phi_l(v) \|_{L_2}
\\ &= \| \sum_{l=1}^{r_n}\hat{\beta}_{jkl}\big( \hat\phi_l(v) - \phi_l(v) \big) + \sum_{l=1}^{r_n} ( \hat\beta_{jkl} -\beta_{jkl0})\phi_{l}(v) - \sum_{l=r_n+1}^{\infty}\beta_{jkl0}\phi_l(v) \|_{L_2}
\\ &\le \| \sum_{l=1}^{r_n}\hat{\beta}_{jkl}\big( \hat\phi_l(v) - \phi_l(v) \big)  \|_{L_2} + \Big\| \sum_{l=1}^{r_n} (\hat{\beta}_{jkl} - \beta_{jkl0})\phi_l(v) \Big\|_{L_2}  +  \Big\| \sum_{l=r_n+1}^{\infty} \beta_{jkl0}\phi_l(v) \Big\|_{L_2} 
\\ & \leq \underbrace{\sum_{l=1}^{r_n} |\hat{\beta}_{jkl}| \cdot \| \hat\phi_l(v) - \phi_l(v) \|_{L_2}}_{\text{(i)}} + \underbrace{\| \hat{\mathcal{B}}_{r_n} - \mathcal{B}_{r_n0} \|}_{\text{(ii)}} + \underbrace{(\sum_{l=r_n+1}^{\infty}\beta_{jkl0}^2)^{1/2}}_{\text{(iii)}}
\\ & = O_p(n^{-1/2}r_n^{a_1 + 3/2}) + O_p(\alpha_n) + O(r_n^{-a_2+1/2})
\\ & = O_p(\alpha_n)
\end{aligned}$$
where term (i) is bounded by applying the Cauchy-Schwarz inequality along with Lemmas~\ref{lemma1} and \ref{lemma2}. Specifically, $\sum_{l=1}^{r_n} |\hat{\beta}_{jkl}| \cdot \| \hat\phi_l(v) - \phi_l(v) \|_{L_2} \le (\sum_{l=1}^{r_n}\hat\beta_{jkl}^2)^{1/2}(\sum_{l=1}^{r_n} \| \hat\phi_l(v) - \phi_l(v) \|_{L_2}^2)^{1/2}$. Since $\sum_{l=1}^{r_n} \hat\beta_{jkl}^2 \leq C$ and 
$\sum_{l=1}^{r_n}\| \hat\phi_l(v) - \phi_l(v) \|_{L_2}^2 \lesssim \hat\Delta^2 \sum_{l=1}^{r_n}\delta_l^{-2}=O_p(r_n^{2a_1+3}n^{-1})$, the convergence rate of term (i) yields $O_p(n^{-1/2}r_n^{a_1 + 3/2})$.
Term (ii) directly follows from the result of Step 2. For term (iii), evaluating the truncation error using Condition~\ref{cond_2:true_param} yields $\sum_{l=r_n+1}^{\infty}\beta_{jkl0}^2 \lesssim \sum_{l=r_n+1}^{\infty} l^{-2a_2} =O(r_n^{-2a_2+1})$. 
Given Condition~\ref{cond_3:truncation_rate} on the rate and Condition~\ref{cond_2:true_param} with $a_2>1$, we observe that the rate of term (i) is the square of the first component of $\alpha_n$, and the rate of term (iii) satisfies $r_n^{-a_2+1/2} \le r_n^{-a_2/2}$ so that both of them are dominated by $O_p(\alpha_n)$.
Consequently, all three terms are bounded by $\alpha_n$, yielding $\|\hat{\beta}_{jk}(v) - \beta_{jk0}(v)\|_{L_2} = O_p(\alpha_n)$, which completes the proof of Theorem 1.

\newpage
\subsection{Proof of Theorem 2}
\label{S1.4:ProofThm2}
Let $\mbf{\zeta} = (\mbf{\theta}, \mathcal{B}_{r_n})$ denote the vector of all finite-dimensional parameters in the truncated working model, and let its dimension be $d_n = |\mathcal{D}|(r_n+d)+1$.
Recall that $\mathcal{H}$ is the nuisance parameter in the profile score test. 
Let $L(\mbf{\zeta}, \mathcal{H})$ and $l(\mbf{\zeta}, \mathcal{H})$ denote the likelihood and log-likelihood functions for a single subject, respectively. 
Let $(\mbf{\zeta}_0, \mathcal{H}_0)$ denote the true values of $(\mbf{\zeta}, \mathcal{H})$ under the null hypothesis $H_0: \mathcal{B}_{r_n} = \mbf{0}$, where $\mbf{\zeta}_0 = (\mbf{\theta}_0, \mbf{0})$. Let $\tilde{\mbf{\zeta}} = (\tilde{\mbf{\theta}}, \mbf{0})$ and $\tilde{\mathcal{H}}$ denote the corresponding restricted maximum pseudo likelihood estimators under $H_0$. 
Let $\tilde{\ell}_{\mbf{\zeta}}(O_i)$ be the efficient score for the joint parameter $\mbf{\zeta}$ based on the $i$th subject's observed data $O_i$, evaluated at $({\mbf{\zeta}}_0, {\mathcal{H}}_0)$, and let $\tilde{\mathcal{I}}_{\mbf{\zeta}} = E\{ \tilde{\ell}_{\mbf{\zeta}}(O_i)^{\otimes 2} \}$ be the efficient information matrix. 
For simplicity, and with a slight abuse of notation, we use $\mbf{\Xi}$ and $\hat{\mbf{\Xi}}$ throughout this proof to denote the true and estimated FPC scores, respectively, both truncated at $r_n$.
Unless explicitly stated otherwise, all quantities in the proof are based on $\mbf{\Xi}$ rather than $\hat{\mbf{\Xi}}$, and all functions are evaluated at the true parameter values $(\mbf{\zeta}_0, \mathcal{H}_0)$. 
Note that under $H_0$, using $\mbf{\Xi}$ or $\hat{\mbf{\Xi}}$ yields the same restricted estimators $(\tilde{\mbf{\zeta}}, \tilde{\mathcal{H}})$.

To formulate the score functions, we first define a building-block integrand for any observation interval $q \in \{1, \ldots, Q\}$, state transition $(j, k) \in \mathcal{D}$, and time $t \in [0, \tau]$:
\begin{equation*}
\begin{aligned}
    B_{qjk}(t; \mbf{\zeta} ,\mathcal{H}) &= \frac{1}{L(\mbf{\zeta}, \mathcal{H})} \int_b \Bigg\{ \prod_{q' \neq q} \mbf{P}(\tau_{q'-1}, \tau_{q'}; \mbf{X}, \mbf{\Xi}, b)^{(S_{q'-1}, S_{q'})} \Bigg\} \\
    &\quad \times \mbf{P}(\tau_{q-1}, t; \mbf{X}, \mbf{\Xi}, b)^{(S_{q-1}, j)} \exp\big\{ \mbf{\gamma}_{jk}^{\T} \mbf{X}(t) + \mbf{B}_{jk}^{\T}\mbf{\Xi} + b \big\} \\
    &\quad \times \Big\{ \mbf{P}(t, \tau_q; \mbf{X}, \mbf{\Xi}, b)^{(k, S_q)} - \mbf{P}(t, \tau_q; \mbf{X}, \mbf{\Xi}, b)^{(j, S_q)} \Big\} \\
    &\quad \times I(\tau_{q-1} < t \leq \tau_q) f(b;\sigma^2) db.
\end{aligned}
\end{equation*}
Accordingly, the score vector for the joint parameter of interest is partitioned as:
\begin{equation*}
    \ell_{\mbf{\zeta}}( \mbf{\zeta} ,\mathcal{H}) = \begin{bmatrix} \ell_{\mbf{\theta}}( \mbf{\zeta} ,\mathcal{H}) \\ \ell_{\mathcal{B}_{r_n}}( \mbf{\zeta} ,\mathcal{H}) \end{bmatrix} = \begin{bmatrix} \ell_{\mbf{\gamma}}( \mbf{\zeta} ,\mathcal{H}) \\ \ell_{\sigma^2}( \mbf{\zeta} ,\mathcal{H}) \\ \ell_{\mathcal{B}_{r_n}}( \mbf{\zeta} ,\mathcal{H}) \end{bmatrix}.
\end{equation*}

Based on the integrand defined above, these score components are explicitly given by:
\begin{align*}
    \ell_{\mbf{\gamma}_{jk}}( \mbf{\zeta} ,\mathcal{H}) &= \sum_{q=1}^{Q} \int_{0}^{\tau} B_{qjk}(t;  \mbf{\zeta} ,\mathcal{H}) \mbf{X}(t) dH_{jk}(t), \\
    \ell_{\mbf{B}_{jk}}( \mbf{\zeta} ,\mathcal{H}) &= \sum_{q=1}^{Q} \int_{0}^{\tau} B_{qjk}(t;  \mbf{\zeta} ,\mathcal{H}) \mbf{\Xi} dH_{jk}(t),
\end{align*}
\begin{equation*}
    \ell_{\sigma^2}(\mbf{\zeta},\mathcal{H}) = \frac{1}{L(\mbf{\zeta} ,\mathcal{H})} \int_b \Bigg\{ \prod_{q=1}^{Q} \mbf{P}(\tau_{q-1}, \tau_q; \mbf{X}, \mbf{\Xi}, b)^{(S_{q-1}, S_q)} \Bigg\} f'(b; \sigma^2) db.
\end{equation*}

The outline of the proof is as follows.
First, we establish the asymptotic null distribution of the efficient score statistic constructed from the efficient score and efficient information for $\mbf{\zeta}$.
Next, we show that this statistic can be well approximated by the profile score statistic based on the true FPC scores $\mbf{\Xi}$.
Finally, we show that replacing $\mbf{\Xi}$ by $\hat{\mbf{\Xi}}$ does not affect the asymptotic null distribution of the profile score statistic.

{\it Step 1.} We establish the asymptotic equivalence between the profile score and the efficient score, where the profile score is defined as $S_n(\tilde{\mbf{\zeta}}) = n\mathbb{P}_n \frac{\partial}{\partial \mbf{\zeta} } l\{\mbf{\zeta},\hat{\mathcal{H}}(\mbf{\zeta})\}\Big|_{\mbf{\zeta} = \tilde{\mbf{\zeta}}}$, with $\hat{\mathcal{H}}(\mbf{\zeta}) = \arg \max_{\mathcal{H}} \mathbb{P}_n l(\mbf{\zeta}, \mathcal{H})$ for a fixed $\mbf{\zeta}$. 
We show that:
\begin{align} \label{step1}
    n^{-1/2}  S_n(\tilde{\mbf{\zeta}}) = n^{-1/2} \sum_{i=1}^n \tilde{\ell}_{\mbf{\zeta}}(O_i) + o_p(1).
\end{align}

To prove this, we first define the collection of functions $\mbf{h}^*(t)= \{\mbf{h}_{jk}^*(t) : (j,k) \in \mathcal{D}\}$, which represents the least favorable direction for $\mathcal{H}$. It is uniquely defined by the operator equation
\begin{equation*}
    \ell_{\mathcal{H}}^* \ell_{\mathcal{H}}[\mbf{h}^*] = \ell_{\mathcal{H}}^* \ell_{\mbf{\zeta}},
\end{equation*}
where $\ell_{\mathcal{H}}$ is the score operator for the nuisance parameter $\mathcal{H}$ and $\ell_{\mathcal{H}}^*$ is the adjoint operator of $\ell_{\mathcal{H}}$. 
Along this least favorable direction, we construct a least favorable submodel $(\mbf{\eta}, \mathcal{H}_{\mbf{\eta}, \mbf{\zeta}})$, where the path for the nuisance parameter is given by $ d\mathcal{H}_{\mbf{\eta},\mbf{\zeta}}(t) = \big\{ 1 +  ( \mbf{\zeta} - \mbf{\eta})^{\T} \mbf{h}^*(t) \big\} d{\mathcal{H}}(t)$ for $\mbf{\eta}$ in a sufficiently small compact neighborhood of $\mbf{\zeta}$. Specifically, this element-wise notation implies $d\mathcal{H}_{jk, \mbf{\eta}, \mbf{\zeta}}(t) = \big\{ 1 + ( \mbf{\zeta} - \mbf{\eta})^{\T} \mbf{h}_{jk}^*(t) \big\} dH_{jk}(t)$ for every $(j,k) \in \mathcal{D}$. For this submodel, we define the first and second pathwise derivatives of the log-likelihood with respect to the perturbation parameter $\mbf{\eta}$ as $\dot{l}(\mbf{\eta}, \mbf{\zeta}, \mathcal{H}) = \frac{\partial}{\partial \mbf{\eta}} l(\mbf{\eta}, \mathcal{H}_{\mbf{\eta},\mbf{\zeta}})$ and $\ddot{l}(\mbf{\eta}, \mbf{\zeta}, \mathcal{H}) = \frac{\partial^2}{\partial \mbf{\eta} \partial \mbf{\eta}^{\T}} l(\mbf{\eta}, \mathcal{H}_{\mbf{\eta},\mbf{\zeta}})$, respectively.


We now formally verify the structural conditions of Theorem 1 in \cite{murphy2000profile}:
\begin{enumerate}
    \item \textbf{Condition (8) (Submodel anchor):} By construction, when evaluating the submodel at $\mbf{\eta} = \mbf{\zeta}$, the perturbation term vanishes, yielding $\mathcal{H}_{\mbf{\eta}, \mbf{\zeta}} = \mathcal{H}$. Thus, the path naturally passes through $(\mbf{\zeta}, \mathcal{H})$.
    
    \item \textbf{Condition (9) (Least favorable direction):} The existence and boundedness of the least favorable direction $\mbf{h}^*$ established in Lemma 3 of \cite{gu2024maximum} directly fulfill this condition, ensuring that the pathwise derivative aligns with the efficient score evaluated at true parameters.
    
    \item \textbf{Condition (10) (Consistency of profiled nuisance parameter):} Theorem 1 of \cite{gu2024maximum} guarantees that the profiled estimator $\hat{\mathcal{H}}(\mbf{\zeta})$ is asymptotically bounded when $\mbf{\zeta}$ is in a small neighborhood of $\mbf{\zeta}_0$, implying that the restricted estimator $\tilde{\mathcal{H}}$ converges to $\mathcal{H}_0$ as $\tilde{\mbf{\zeta}} \to \mbf{\zeta}_0$. Furthermore, by the arguments of Lemma 2 in \cite{gu2024maximum}, we obtain the uniform convergence rate $\hat{\mathcal{H}}(\mbf{\zeta}) = \mathcal{H}_0 + O_p(\|\mbf{\zeta} - \mbf{\zeta}_0\|) + O_p(n^{-1/3})$, which satisfies the consistency requirement. 
    
    \item \textbf{Condition (11) (No-bias condition):} Crucially, the $O_p(n^{-1/3})$ convergence rate established above is strictly faster than $o_p(n^{-1/4})$. According to the general verification scheme outlined in \cite{murphy2000profile}, this fast convergence ensures that the second-order Taylor expansion of the expected score under the true probability measure $\mathbb{P}\dot{l}(\mbf{\zeta}_0, \tilde{\mbf{\zeta}}, \tilde{\mathcal{H}})$ vanishes sufficiently fast, thereby strictly satisfying the asymptotic no-bias requirement.
\end{enumerate}


Finally, we verify the empirical process regularity conditions required by Theorem 1 of \cite{murphy2000profile}, which require that the first and second pathwise derivatives of the log-likelihood along the submodel must belong to a $P$-Donsker class. By Lemma 1 in \cite{gu2024maximum}, the score vector $\ell_{\mbf{\zeta}}(\mbf{\zeta}, \mathcal{H})$ and the score operator $\ell_{\mathcal{H}}(\mbf{\zeta}, \mathcal{H})[\mbf{h}^*]$ are already established as $P$-Donsker. 

Applying the chain rule, the first pathwise derivative with respect to $\mbf{\eta}$ naturally emerges as a linear combination of these Donsker components:
\begin{equation*}
    \dot{l}(\mbf{\eta}, \mbf{\zeta} ,\mathcal{H}) = \ell_{\mbf{\zeta}}(\mbf{\zeta}, \mathcal{H}_{\mbf{\eta},\mbf{\zeta}}) - \ell_{\mathcal{H}}(\mbf{\zeta}, \mathcal{H}_{\mbf{\eta},\mbf{\zeta}})[\mbf{h}^*].
\end{equation*}
Because the perturbation $\mbf{\eta}$ is constrained to a compact neighborhood and the submodel path $\mathcal{H}_{\mbf{\eta},\mbf{\zeta}}$ is sufficiently smooth with respect to $(\mbf{\eta}, \mbf{\zeta})$, this composite derivative $\dot{l}(\mbf{\eta}, \mbf{\zeta} ,\mathcal{H})$ safely preserves the Donsker property. By an identical smoothness argument, the second pathwise derivative $\ddot{l}(\mbf{\eta}, \mbf{\zeta} , \mathcal{H})$ also forms a Donsker class. 

Consequently, all regularity conditions are verified, allowing us to invoke Theorem 1 of \cite{murphy2000profile} to establish the asymptotic expansion of the profile likelihood and conclude equation \eqref{step1}.

{\it Step 2.} Under \(H_0\), the efficient scores \(\tilde l_{\mbf{\zeta}}(O_i),i=1,\ldots,n\) are i.i.d. random variables with \(E\{\tilde l_{\mbf{\zeta}}(O_i)\} = 0\).
Define $\mbf{\psi}_i = \tilde{\mathcal{I}}_{\mbf{\zeta}}^{-1/2} \tilde l_{\mbf{\zeta}}(O_i) $ and \(\tilde{\mbf{\psi}}_n = n^{-1/2} \sum_{i=1}^n \mbf{\psi}_i\). The \(\mbf{\psi}_i\)'s are i.i.d. with mean zero and covariance \(\mbf{I}_{d_n}\). Under Conditions \ref{cond_4:functional_M1} and \ref{cond6:X}, the sub-exponential tail of the functional covariate $\|M\|_{L_2}$ and the boundedness of the scalar covariates guarantee that each component of the efficient score vector $\tilde{\ell}_{\mbf{\zeta}}(O_i)$ possesses a uniformly bounded fourth moment, independent of the dimension $d_n$. Consequently, expanding the squared Euclidean norm yields $E\Big[ \big\{ \tilde{\ell}_{\mbf{\zeta}}(O_i)^{\T} \tilde{\ell}_{\mbf{\zeta}}(O_i) \big\}^2 \Big] = O(d_n^2)$.
Furthermore, the divergence rate $r_n^{2a_1+3} = o(n)$ imposed in Condition \ref{cond_3:truncation_rate} implies $d_n = o(n)$, which ensures $O(d_n^2) = o(n d_n)$. Thus, we obtain $E\{ (\tilde l_{\mbf{\zeta}}(O_i)^{\T} \tilde l_{\mbf{\zeta}}(O_i))^2 \} = O(d_n^2) = o(nd_n)$. By invoking Lemma~\ref{lemma3}, this yields:
$$\frac{(\tilde{\mbf{\psi}}_n^{\T} \tilde{\mbf{\psi}}_n - d_n)}{(2 d_n)^{1/2}} = \frac{\left\{n^{-1/2}\sum_{i=1}^{n}\tilde{\mathcal{{I}}}_{\mbf{\zeta}}^{-1/2}\tilde{l}_{\mbf{\zeta}}(O_{i})\right\}^{\T}\left\{n^{-1/2}\sum_{i=1}^{n}\tilde{\mathcal{{I}}}_{\mbf{\zeta}}^{-1/2}\tilde{l}_{\mbf{\zeta}}(O_{i})\right\}-d_{n}}{\left(2d_{n}\right)^{1/2}} \overset{d}{\to} N(0,1)$$
Using result \eqref{step1}, we obtain:
$$ \frac{\left\{n^{-1/2}\tilde{\mathcal{{I}}}_{\mbf{\zeta}}^{-1/2}S_n(\tilde{\mbf{\zeta}})\right\}^{\T}\left\{n^{-1/2}\tilde{\mathcal{{I}}}_{\mbf{\zeta}}^{-1/2}S_n(\tilde{\mbf{\zeta}})\right\}
-d_{n}}{\left(2d_{n}\right)^{1/2}} = \frac{n^{-1}S_n(\tilde{\mbf{\zeta}})^{\T}\tilde{\mathcal{I}}_{\mbf{\zeta}}^{-1}S_n(\tilde{\mbf{\zeta}})
-d_{n}}{\left(2d_{n}\right)^{1/2}}
\overset{d}{\to}N(0,1)$$

{\it Step 3.} To estimate the covariance matrix, we rely on the empirical information matrix of the profile log-likelihood evaluated at the restricted estimator $\tilde{\mbf{\zeta}} = (\tilde{\mbf{\theta}}, \mbf{0})$. The empirical efficient information matrix with respect to  $\mbf{\zeta}$ is defined as:

\begin{align*}
    \hat{\mbf{V}}_n &= \mathbb{P}_n\left[\left.\frac{\partial}{\partial \mbf{\zeta}}l\{\mbf{\zeta},\hat{\mathcal{H}}(\mbf{\zeta})\}\right|_{\mbf{\zeta}=\tilde{\mbf{\zeta}}}\right]^{\otimes2}
    \\ & = \mathbb{P}_n\left[ \left. \ell_{\mbf{\zeta}}\{\mbf{\zeta},\hat{\mathcal{H}}(\mbf{\zeta})\}  \right|_{\mbf{\zeta}=\tilde{\mbf{\zeta}}} + \left. \ell_{\mathcal{H}}\{\mbf{\zeta},\hat{\mathcal{H}}(\mbf{\zeta})\}\left[\frac{\partial\hat{\mathcal{H}}(\mbf{\zeta})}{\partial\mbf{\zeta}}\right] \right|_{\mbf{\zeta}=\tilde{\mbf{\zeta}}} \right]^{\otimes2}
    \\ & = \mathbb{P}_n\Bigg[\ell_{\mbf{\zeta}}\{\tilde{\mbf{\zeta}},\hat{\mathcal{H}}(\tilde{\mbf{\zeta}})\}   + \ell_{\mathcal{H}}\{\tilde{\mbf{\zeta}},\hat{\mathcal{H}}(\tilde{\mbf{\zeta}})\}\left[ \dot{\mathcal{H}}_{\tilde{\mbf{\zeta}}} \right] \Bigg]^{\otimes2},
\end{align*}
where the second equality is obtained by applying the chain rule and $\dot{\mathcal{H}}_{\tilde{\mbf{\zeta}}} = \frac{\partial\hat{\mathcal{H}}(\mbf{\zeta})}{\partial\mbf{\zeta}} \big|_{\mbf{\zeta}=\tilde{\mbf{\zeta}}}$.
We aim to prove that $\hat{\mbf{V}}_n - \tilde{\mathcal{I}}_{\mbf{\zeta}} = o_p(1)$.


By the definition of $\hat{\mathcal{H}}(\mbf{\zeta})$, the score equation $\mathbb{P}_n \ell_{\mathcal{H}} \{\mbf{\zeta},\hat{\mathcal{H}}(\mbf{\zeta})\}[h] = 0$ holds for any given finite-dimensional parameter $\mbf{\zeta}$ and direction $h \in \mathbb{H} = \mathrm{BV}[0,\tau]^{|\mathcal{D}|}$, where $\mathrm{BV}[0,\tau]$ denotes the set of all functions with bounded total variation over $[0,\tau]$. Differentiating this equation with respect to $\mbf{\zeta}$ at $\mbf{\zeta} = \tilde{\mbf{\zeta}}$ yields
\begin{equation} \label{eq:score_diff}
    \mathbb{P}_n \ell_{\mathcal{H}\mathcal{H}} (\tilde{\mbf{\zeta}},\tilde{\mathcal{H}})[h, \dot{\mathcal{H}}_{\tilde{\mbf{\zeta}}}] = - \mathbb{P}_n \ell_{\mathcal{H}\mbf{\zeta}} (\tilde{\mbf{\zeta}},\tilde{\mathcal{H}})[h], \quad \text{for any } h \in \mathbb{H}.
\end{equation}
where $\ell_{\mathcal{H}\mbf{\zeta}} (\tilde{\mbf{\zeta}},\tilde{\mathcal{H}})[h]$ is the derivative of $\ell_{\mathcal{H}}(\tilde{\mbf{\zeta}},\tilde{\mathcal{H}})[h]$ with respect to $\mbf{\zeta}$ and $\ell_{\mathcal{H}\mathcal{H}} (\tilde{\mbf{\zeta}},\tilde{\mathcal{H}})[h, \dot{\mathcal{H}}_{\tilde{\mbf{\zeta}}}]$ is the derivative of $\ell_{\mathcal{H}}(\tilde{\mbf{\zeta}},\tilde{\mathcal{H}})[h]$ with respect to $\mathcal{H}$ along the direction $\dot{\mathcal{H}}_{\tilde{\mbf{\zeta}}}$.

Since the theoretical operator $\mathbb{P} \ell_{\mathcal{H}\mathcal{H}} (\mbf{\zeta}_0,\mathcal{H}_0)$ is invertible and $\| \mathbb{P}_n \ell_{\mathcal{H}\mathcal{H}} (\tilde{\mbf{\zeta}},\tilde{\mathcal{H}}) - \mathbb{P} \ell_{\mathcal{H}\mathcal{H}} (\mbf{\zeta}_0,\mathcal{H}_0)\| \to 0$ in probability, the empirical operator $\mathbb{P}_n \ell_{\mathcal{H}\mathcal{H}} (\tilde{\mbf{\zeta}},\tilde{\mathcal{H}})$ is invertible for sufficiently large $n$. This invertibility guarantees that the empirical score equation \eqref{eq:score_diff} yields a unique solution $\dot{\mathcal{H}}_{\tilde{\mbf{\zeta}}}$, which possesses bounded total variation. Consequently, this bounded total variation property ensures that the operator $\ell_{\mathcal{H}}(\tilde{\mbf{\zeta}},\tilde{\mathcal{H}})[\dot{\mathcal{H}}_{\tilde{\mbf{\zeta}}}]$ belongs to a Donsker class. 
Furthermore, following similar arguments to Lemma S.3 of \citet{gu2025semiparametric}, we can establish $-\mathbb{P}\ell_{\mathcal{H}\mathcal{H}}(\mbf{\zeta}_0,\mathcal{H}_0)[\dot{\mathcal{H}}_{\tilde{\mbf{\zeta}}} +\mbf{h}^*, \dot{\mathcal{H}}_{\tilde{\mbf{\zeta}}} +\mbf{h}^*] \ge C\|\dot{\mathcal{H}}_{\tilde{\mbf{\zeta}}}+\mbf{h}^*\|_{L_2}^2$ for some positive constant $C$. By the Donsker property, we can conclude that the empirical counterpart $-\mathbb{P}_n\ell_{\mathcal{H}\mathcal{H}}(\tilde{\mbf{\zeta}},\tilde{\mathcal{H}})[\dot{\mathcal{H}}_{\tilde{\mbf{\zeta}}} +\mbf{h}^*, \dot{\mathcal{H}}_{\tilde{\mbf{\zeta}}} +\mbf{h}^*] \ge C\|\dot{\mathcal{H}}_{\tilde{\mbf{\zeta}}}+\mbf{h}^*\|_{L_2}^2$ holds for sufficiently large $n$.

Substituting the specific direction $h = \dot{\mathcal{H}}_{\tilde{\mbf{\zeta}}} + \mbf{h}^*$ into equation \eqref{eq:score_diff}, 
\begin{align*}
-\mathbb{P}_{n}\ell_{\mathcal{H}\mathcal{H}}(\tilde{\mbf{\zeta}},\tilde{\mathcal{H}})[ \dot{\mathcal{H}}_{\tilde{\mbf{\zeta}}}+\mbf{h}^{*}, \dot{\mathcal{H}}_{\tilde{\mbf{\zeta}}}] &= \mathbb{P}_{n}\ell_{\mbf{\zeta}\mathcal{H}}(\tilde{\mbf{\zeta}},\tilde{\mathcal{H}})[\dot{\mathcal{H}}_{\tilde{\mbf{\zeta}}}+\mbf{h}^{*}] \\
&= o_{p}(1)\|\dot{\mathcal{H}}_{\tilde{\mbf{\zeta}}}+\mbf{h}^{*}\|_{L_{2}} +\mathbb{P}\ell_{\mbf{\zeta}\mathcal{H}}(\mbf{\zeta}_{0},\mathcal{H}_{0})[\dot{\mathcal{H}}_{\tilde{\mbf{\zeta}}}+\mbf{h}^{*}],
\end{align*}
and similarly,
\begin{align*}
-\mathbb{P}_n\ell_{\mathcal{H}\mathcal{H}}(\tilde{\mbf{\zeta}},\tilde{\mathcal{H}})[\dot{\mathcal{H}}_{\tilde{\mbf{\zeta}}}+\mbf{h}^*, \mbf{h}^*] &= -\mathbb{P}\ell_{\mathcal{H}\mathcal{H}}(\mbf{\zeta}_0,\mathcal{H}_0)[\dot{\mathcal{H}}_{\tilde{\mbf{\zeta}}}+\mbf{h}^*, \mbf{h}^*] + o_{p}(1)\|\dot{\mathcal{H}}_{\tilde{\mbf{\zeta}}}+\mbf{h}^*\|_{L_2} \\
&= -\mathbb{P}\ell_{\mbf{\zeta}\mathcal{H}}(\mbf{\zeta}_0,\mathcal{H}_0)[\dot{\mathcal{H}}_{\tilde{\mbf{\zeta}}}+\mbf{h}^*] + o_{p}(1)\|\dot{\mathcal{H}}_{\tilde{\mbf{\zeta}}}+\mbf{h}^*\|_{L_2}.
\end{align*}
Taking the sum yields:
$$
-\mathbb{P}_n\ell_{\mathcal{H}\mathcal{H}}(\tilde{\mbf{\zeta}},\tilde{\mathcal{H}})[\dot{\mathcal{H}}_{\tilde{\mbf{\zeta}}} +\mbf{h}^*, \dot{\mathcal{H}}_{\tilde{\mbf{\zeta}}} +\mbf{h}^*] \leq o_{p}(1)\|\dot{\mathcal{H}}_{\tilde{\mbf{\zeta}}}+\mbf{h}^*\|_{L_2}.
$$
Combined with the previous result $-\mathbb{P}_n\ell_{\mathcal{H}\mathcal{H}}(\tilde{\mbf{\zeta}},\tilde{\mathcal{H}})[\dot{\mathcal{H}}_{\tilde{\mbf{\zeta}}} +\mbf{h}^*, \dot{\mathcal{H}}_{\tilde{\mbf{\zeta}}} +\mbf{h}^*] \ge C\|\dot{\mathcal{H}}_{\tilde{\mbf{\zeta}}}+\mbf{h}^*\|_{L_2}^2$, this implies $C\|\dot{\mathcal{H}}_{\tilde{\mbf{\zeta}}}+\mbf{h}^*\|_{L_2}^2 \leq o_{p}(1)\|\dot{\mathcal{H}}_{\tilde{\mbf{\zeta}}}+\mbf{h}^*\|_{L_2}$. Thus, we obtain $\|\dot{\mathcal{H}}_{\tilde{\mbf{\zeta}}}+\mbf{h}^*\|_{L_2} = o_{p}(1)$.

By the Donsker property of the score operators and the consistency of $(\tilde{\mbf{\zeta}}, \tilde{\mathcal{H}})$, we apply the continuous mapping theorem to obtain:
\begin{align*}
    \hat{\mbf{V}}_n &= \mathbb{P}\bigg[\big\{\ell_{\mbf{\zeta}}(\tilde{\mbf{\zeta}},\tilde{\mathcal{H}}) + \ell_{\mathcal{H}}(\tilde{\mbf{\zeta}},\tilde{\mathcal{H}})[\dot{\mathcal{H}}_{\tilde{\mbf{\zeta}}}]\big\}^{\otimes2}\bigg] + o_{p}(1) \\
    &= \mathbb{P}\bigg[\big\{\ell_{\mbf{\zeta}}({\mbf{\zeta}_{0}},{\mathcal{H}_{0}}) - \ell_{\mathcal{H}}({\mbf{\zeta}_{0}},{\mathcal{H}_{0}})[\mbf{h}^*]\big\}^{\otimes2}\bigg] + o_{p}(1) = \tilde{\mathcal{I}}_{\mbf{\zeta}} + o_{p}(1).
\end{align*}

Consequently, we obtain:
$$
\frac{n^{-1}S_n(\tilde{\mbf{\zeta}})^{\T}\hat{V}_n^{-1}S_n(\tilde{\mbf{\zeta}})
-d_{n}}{\left(2d_{n}\right)^{1/2}}
\overset{d}{\to}N(0,1)
$$

\textit{Step 4.} We consider the difference of replacing the true latent FPC scores $\mbf{\Xi}$ with their estimates $\hat{\mbf{\Xi}}$. 
We write the profile score $S_n(\tilde{\mbf{\zeta}})$ and the empirical information matrix $\hat{\mbf{V}}_n$ defined in previous steps as $S_n(\tilde{\mbf{\zeta}}; \mbf{\Xi})$ and $\hat{\mbf{V}}_n(\mbf{\Xi})$, respectively, to emphasize that they are both based on the true FPC scores $\mbf{\Xi}$. 
Let $S_n(\tilde{\mbf{\zeta}}; \hat{\mbf{\Xi}})$ and $\hat{\mbf{V}}_n(\hat{\mbf{\Xi}})$ denote the corresponding estimators evaluated at $\hat{\mbf{\Xi}}$. We will show that the differences are asymptotically negligible:
$$
n^{-1/2} \big\| S_n(\tilde{\mbf{\zeta}}; \hat{\mbf{\Xi}}) - S_n(\tilde{\mbf{\zeta}}; \mbf{\Xi}) \big\| = o_p(1), \quad n^{-1} \big\| \hat{\mbf{V}}_n(\hat{\mbf{\Xi}}) - \hat{\mbf{V}}_n(\mbf{\Xi}) \big\| = o_p(1).
$$
The score difference can be decomposed into three terms:
\begin{align}
    n^{-1/2} \big\{ S_n(\tilde{\mbf{\zeta}}; \hat{\mbf{\Xi}}) &- S_n(\tilde{\mbf{\zeta}}; \mbf{\Xi}) \big\} \notag \\
    &= n^{-1/2} \bigg\{ S_n(\tilde{\mbf{\zeta}}; \hat{\mbf{\Xi}}) - \sum_{i=1}^{n}\tilde{\ell}_{\mbf{\zeta}}(O_i; \hat{\mbf{\Xi}}) \bigg\} \label{SD 1} \\
    &\quad + n^{-1/2} \sum_{i=1}^{n} \big\{ \tilde{\ell}_{\mbf{\zeta}}(O_i; \hat{\mbf{\Xi}}) - \tilde{\ell}_{\mbf{\zeta}}(O_i; \mbf{\Xi}) \big\} \label{SD 2} \\
    &\quad + n^{-1/2} \bigg\{ \sum_{i=1}^{n}\tilde{\ell}_{\mbf{\zeta}}(O_i; \mbf{\Xi}) - S_n(\tilde{\mbf{\zeta}}; \mbf{\Xi}) \bigg\}. \label{SD 3}
\end{align}
By Theorem 1 of \cite{murphy2000profile} and the results established in Step 1, both \eqref{SD 1} and \eqref{SD 3} are $o_p(1)$. 

Now, we focus on bounding \eqref{SD 2}. By the definition of the efficient score $\tilde{\ell}_{\mbf{\zeta}} = \ell_{\mbf{\zeta}} - \ell_{\mathcal{H}}[\mbf{h}^*]$, its Euclidean norm is bounded by the triangle inequality:
\begin{align*}
    \| \eqref{SD 2} \| &= \bigg\| n^{-1/2}\sum_{i=1}^{n} \Big\{ \ell_{\mbf{\zeta}}(O_{i}; \hat{\mbf{\Xi}}) - \ell_{\mathcal{H}}(O_{i}; \hat{\mbf{\Xi}})[\mbf{h}^{*}] - \ell_{\mbf{\zeta}}(O_{i}; \mbf{\Xi}) + \ell_{\mathcal{H}}(O_{i}; \mbf{\Xi})[\mbf{h}^{*}] \Big\} \bigg\| \\
    &\leq n^{-1/2}\sum_{i=1}^{n} \Big\{ \big\| \ell_{\mbf{\zeta}}(O_{i}; \hat{\mbf{\Xi}}) - \ell_{\mbf{\zeta}}(O_{i}; \mbf{\Xi}) \big\| + \big\| \ell_{\mathcal{H}}(O_{i}; \hat{\mbf{\Xi}})[\mbf{h}^{*}] - \ell_{\mathcal{H}}(O_{i}; \mbf{\Xi})[\mbf{h}^{*}] \big\| \Big\}.
\end{align*} 

First, we study the difference for the finite-dimensional score vector $\ell_{\mbf{\zeta}} = [\ell_{\mbf{\gamma}}^{\T}, \ell_{\sigma^2}, \ell_{\mathcal{B}_{r_n}}^{\T}]^{\T}$. For the functional effects components, corresponding to each $(j,k) \in \mathcal{D}$, we have:
\begin{align*}
    \big\| \ell_{\mbf{B}_{jk}}(O_i; \hat{\mbf{\Xi}}) - \ell_{\mbf{B}_{jk}}(O_i; \mbf{\Xi}) \big\| &= \bigg\| \sum_{q=1}^{Q}\int_{0}^{\tau} \big[ B_{qjk}(t; \hat{\mbf{\Xi}}) \hat{\mbf{\Xi}} - B_{qjk}(t; \mbf{\Xi}) \mbf{\Xi} \big] dH_{jk}(t) \bigg\| \\
    &\leq \sum_{q=1}^{Q}\int_{0}^{\tau} \Big\{ \big| B_{qjk}(t; \hat{\mbf{\Xi}}) - B_{qjk}(t; \mbf{\Xi}) \big| \|\hat{\mbf{\Xi}}\| + \big| B_{qjk}(t; \mbf{\Xi}) \big| \|\hat{\mbf{\Xi}} - \mbf{\Xi}\| \Big\} dH_{jk}(t).
\end{align*}
Following the arguments of the proof of Lemma 1 in \cite{gu2024maximum}, the Lipschitz property of the transition probabilities yields:
$$
\big| \mbf{P}(t_1,t_2; \hat{\mbf{\Xi}})^{(j,k)} - \mbf{P}(t_1,t_2; \mbf{\Xi})^{(j,k)} \big| \leq C \|\hat{\mbf{\Xi}} - \mbf{\Xi}\|.
$$ 
Applying this to the building-block integrand, we obtain $|B_{qjk}(t; \hat{\mbf{\Xi}}) - B_{qjk}(t; \mbf{\Xi})| \leq C \|\hat{\mbf{\Xi}} - \mbf{\Xi}\|$. Since $|B_{qjk}(t; \mbf{\Xi})|$ is bounded and the cumulative hazard $H_{jk}(t)$ has bounded total variation, integrating over $[0, \tau]$ easily yields:
$$
\big\| \ell_{\mbf{B}_{jk}}(O_i; \hat{\mbf{\Xi}}) - \ell_{\mbf{B}_{jk}}(O_i; \mbf{\Xi}) \big\| \leq O_p(\|\hat{\mbf{\Xi}} - \mbf{\Xi}\|) .
$$
Similar bounding arguments apply to the finite-dimensional components, giving $\| \ell_{\mbf{\gamma}}(O_i; \hat{\mbf{\Xi}}) - \ell_{\mbf{\gamma}}(O_i; \mbf{\Xi}) \| \leq O_p( \|\hat{\mbf{\Xi}} - \mbf{\Xi}\|)$ and $| \ell_{\sigma^2}(O_i; \hat{\mbf{\Xi}}) - \ell_{\sigma^2}(O_i; \mbf{\Xi}) | \leq O_p(\|\hat{\mbf{\Xi}} - \mbf{\Xi}\|)$.

Next, we study the difference between the score function for $\mathcal{H}$ along the least favorable direction $\mbf{h}^*$. Note that by Condition~\ref{cond_2:true_param} and the proof of Lemma 3 of \citet{gu2024maximum}, $\mbf{h}^*$ and $H_{0jk}(t)$ have bounded total variation in $[0, \tau]$. Thus:
\begin{align*}
    \big\| \ell_{\mathcal{H}}(O_{i}; \hat{\mbf{\Xi}})[\mbf{h}^{*}] - \ell_{\mathcal{H}}(O_{i}; \mbf{\Xi})[\mbf{h}^{*}] \big\| &= \bigg\| \sum_{(j,k)\in\mathcal{D}} \sum_{q=1}^{Q} \int_{0}^{\tau} \big[ B_{qjk}(t; \hat{\mbf{\Xi}}) - B_{qjk}(t; \mbf{\Xi}) \big] \mbf{h}_{jk}^*(t) dH_{0jk}(t) \bigg\| \\
    &\leq C \|\hat{\mbf{\Xi}} - \mbf{\Xi}\| = O_p(\|\hat{\mbf{\Xi}} - \mbf{\Xi}\|).
\end{align*}
Combining these bounds, the total difference \eqref{SD 2} is stochastically bounded by $O_p(\|\hat{\mbf{\Xi}} - \mbf{\Xi}\|)$.
Given the convergence rate of the FPC scores derived in Theorem 1, $\|\hat{\mbf{\Xi}} - \mbf{\Xi}\| = O_p(r_n^{a_1+3/2} n^{-1/2}) = o_p(1)$ by Chebyshev inequality and Condition~\ref{cond_3:truncation_rate}. Therefore, the score difference satisfies $n^{-1/2} \| S_n(\tilde{\mbf{\zeta}}; \hat{\mbf{\Xi}}) - S_n(\tilde{\mbf{\zeta}}; \mbf{\Xi}) \| = o_p(1)$.

By an analogous argument exploiting the Lipschitz continuity of the second-order derivatives, we can establish that the empirical information matrix difference $n^{-1} \| \hat{\mbf{V}}_n(\hat{\mbf{\Xi}}) - \hat{\mbf{V}}_n(\mbf{\Xi}) \| = o_p(1)$. 

Consequently, we obtain:
$$
\frac{n^{-1}S_n(\tilde{\mbf{\zeta}}; \mbf{\hat{\Xi}})^{\T}\hat{\mbf{V}}_n(\hat{\Xi})^{-1}S_n(\tilde{\mbf{\zeta}}; \hat{\mbf{\Xi}})
-d_{n}}{\left(2d_{n}\right)^{1/2}}
\overset{d}{\to}N(0,1)
$$

\textit{Step 5.} We bridge the asymptotic results of the joint parameter $\mbf{\zeta}$ to the proposed test statistic $T_n$ for the functional sub-vector $\mathcal{B}_{r_n}$. By the definition of the restricted estimator $\tilde{\mbf{\zeta}} = (\tilde{\mbf{\theta}}, \mbf{0})$, $\tilde{\mbf{\theta}}$ maximizes the profile pseudo-likelihood under the null hypothesis $H_0$. Consequently, the $\mbf{\theta}$-components of the profile score vector $S_n(\tilde{\mbf{\zeta}}; \hat{\mbf{\Xi}})$ are exactly zero. The remaining non-zero block corresponds exclusively to the functional effects, which is defined as the vector $\mbf{U}_n(\tilde{\mbf{\theta}}, \mbf{0})$ evaluated at $\hat{\mbf{\Xi}}$. 
By the standard property of block matrix inversion, the quadratic form of the joint score vector with the inverse of the joint empirical information matrix $\hat{\mbf{V}}_n(\hat{\mbf{\Xi}})^{-1}$ algebraically reduces to the quadratic form of $\mbf{U}_n(\tilde{\mbf{\theta}}, \mbf{0})$ with the inverse of the empirical Schur complement $\mathcal{I}_n(\tilde{\mbf{\theta}}, \mbf{0}; \hat{\mbf{\Xi}})^{-1}$. 
Therefore, the test statistic for the joint parameter reduces to the plug-in score test statistic $T_n$ for the functional subset, maintaining the asymptotic behavior. We conclude the proof of Theorem 2:
$$
\frac{ \mbf{U}_n(\tilde{\mbf{\theta}}, \mbf{0})^{\T} \mathcal{I}_n(\tilde{\mbf{\theta}}, \mbf{0})^{-1} \mbf{U}_n(\tilde{\mbf{\theta}}, \mbf{0}) - |\mathcal{D}|r_n}{\left(2|\mathcal{D}|r_n\right)^{1/2}} \stackrel{d}{\to} N(0,1).
$$

\bibliographystyle{imsart-nameyear} 
\bibliography{paper-ref}       

\clearpage


\begin{figure}[!t]
\centering
\includegraphics[width=\linewidth]{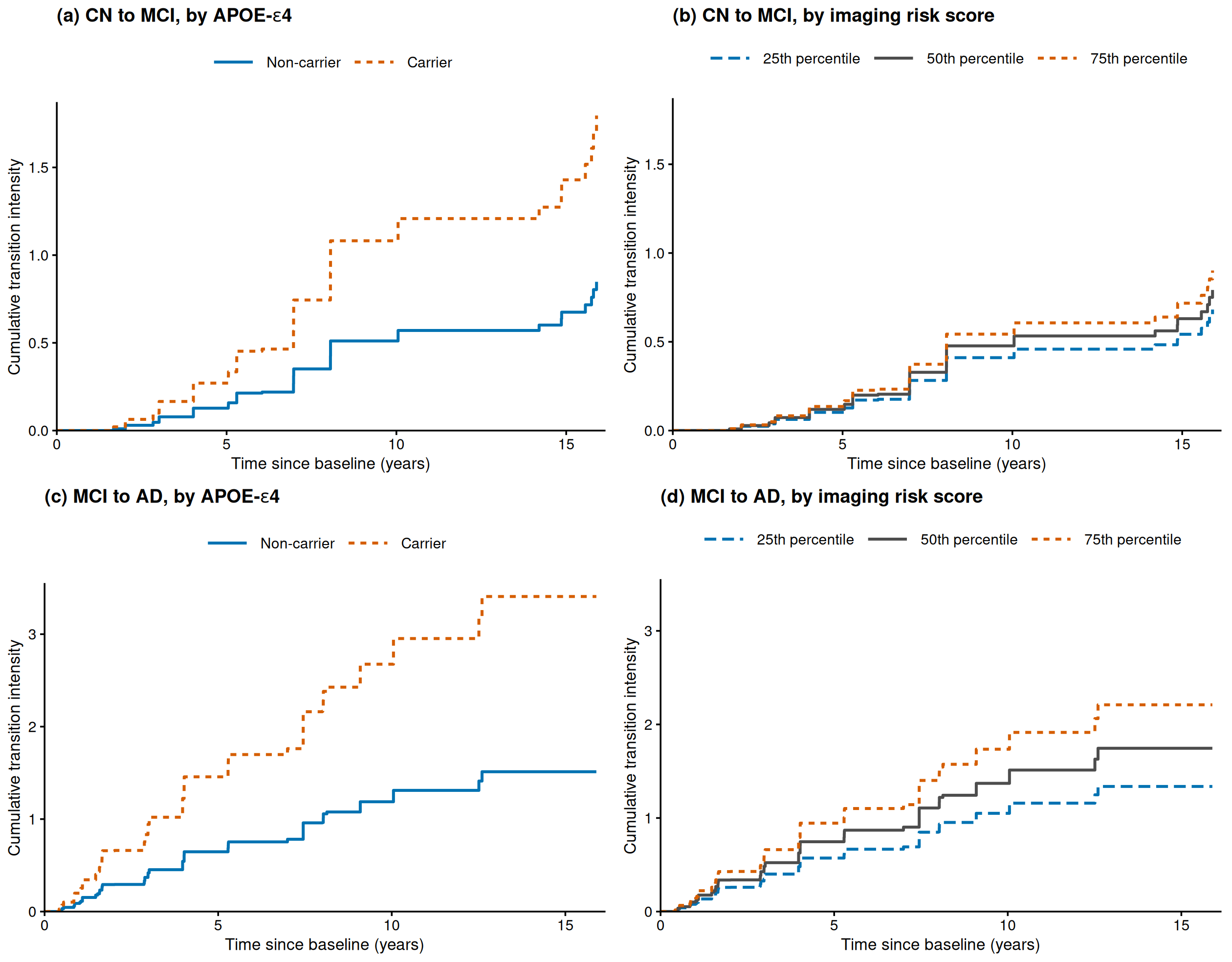}
\caption{Estimated cumulative transition intensity functions in the ADNI cohort under the model with $r_n=4$ (99\% PVE). The left panels are stratified by APOE-$\varepsilon4$ carrier status, and the right panels are stratified by the 25th, 50th, and 75th percentiles of the corresponding transition-specific imaging risk score. The top row shows the CN-to-MCI
transition, and the bottom row shows the MCI-to-AD transition. All remaining covariates are fixed at their sample medians.}
\label{fig:cumulative_hazard_apoe4_rn4}
\end{figure}

\begin{figure}
    \centering
    \includegraphics[width=0.6\linewidth]{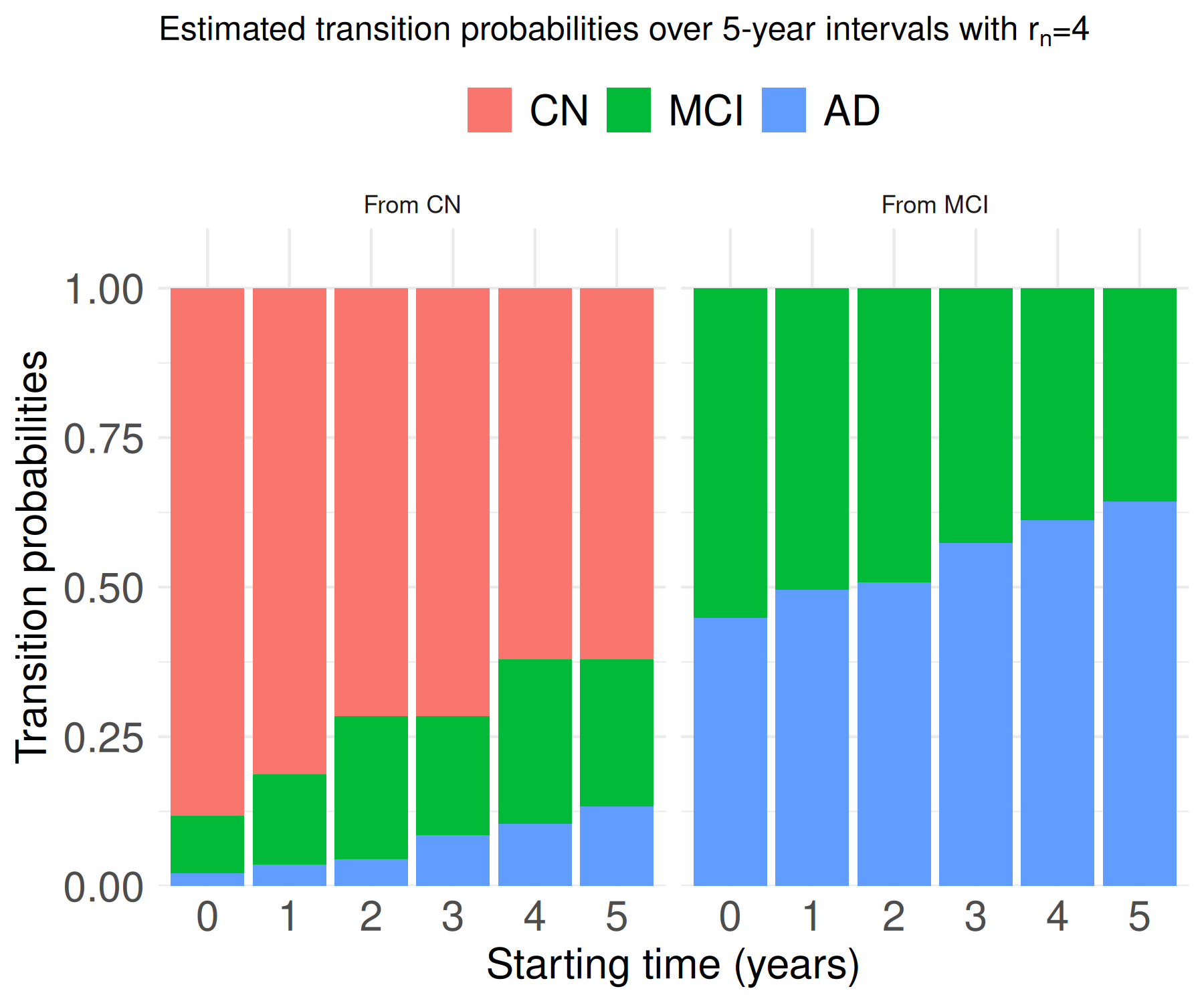}
    \caption{Estimated transition probabilities over five-year time intervals for a subject with baseline covariates fixed at their sample medians under the fitted functional multistate model with $r_n=4$ (99\% PVE).}
    \label{fig:ADNI_trans_prob_rn4}
\end{figure}

\label{lastpage}